\documentclass[%
 reprint,
 amsmath,amssymb,
 aps,
 longbibliography,
 pre,
]{revtex4-1}

\usepackage{graphicx}
\usepackage{dcolumn}
\usepackage{bm}
\usepackage[mathlines]{lineno}
\usepackage{cleveref}
\usepackage{color}

\usepackage[titletoc,title]{appendix}

\AtBeginDocument{%
    \newwrite\bibnotes
    \def\bibnotesext{Notes.bib}
 \immediate\openout\bibnotes=\jobname\bibnotesext
  \immediate\write\bibnotes{@CONTROL{REVTEX41Control}}
    \immediate\write\bibnotes{@CONTROL{%
    apsrev41Control,author="08",editor="1",pages="1",title="0",year="1"}}
     \if@filesw
  \immediate\write\@auxout{\string\citation{apsrev41Control}}%
    \fi
}%

\begin{document}

\preprint{APS/123-QED}

\title{Resilience in Hierarchical Fluid Flow Networks}

\author{Tatyana Gavrilchenko}
\email{tatyanag@sas.upenn.edu}
\author{Eleni Katifori}%
\affiliation{%
Department of Physics and Astronomy, University of Pennsylvania
}%

\date{\today}

\begin{abstract}
The structure of flow networks determines their  function under normal conditions as well as their response to perturbative damage. Brain vasculature often experiences transient or permanent occlusions in the finest vessels, but it is not clear how these micro-clots affect the large scale blood flow or to what extent they decrease functionality. Motivated by this, we investigate how flow is rerouted after the occlusion of a single edge in networks with a hierarchy in edge conductances. We find that in 2D networks, vessels formed by highly conductive edges serve as barriers to contain the displacement of flow due to a localized perturbation. In this way, the vein provides shielding from damage to surrounding edges. We show that once the conductance of the vein surpasses an initial minimal value, further increasing the conductance can no longer extend the shielding provided by the vein. Rather, the length scale of the shielding is set by the network topology. Upon understanding the effects of a single vein, we investigate the global resilience of networks with complex hierarchical order. We find that a system of veins arranged in a grid is able to modestly increase the overall network resilience, outperforming a parallel vein pattern.
\end{abstract}

\maketitle

\section{Introduction}

Damage and recovery play an important role in how  biological and man-made flow networks are designed and operate. Depending on the network architecture, it is possible to inflict massive cascading failure in a functioning network by knocking out just a few key nodes or edges \cite{Watts2002}. Previous work on power grid networks has sought to identify vulnerable edges that are most susceptible to overload and will cause global failure if removed \cite{Witthaut2016, Manik2017, Albert2004, Tyloo2018}. Another example arises in ecological networks, where removing a keystone species can result in the collapse of an ecosystem \cite{Sole2001}. However, in many cases a network is able to sustain damage without complete failure. Recent work on network structure has identified architectural and topological features that allow networks to withstand limited damage or operate in unstable fluctuating conditions \cite{Kaluza2007, Hu2016, Gao2016}. For example, architectures with many hierarchically nested loops allow complex networks to maintain optimal function in the presence of load fluctuations or damage \cite{Banavar2000, Katifori2010, Corson2010}, and the wiring of scale-free networks increases  tolerance to random failures and renders the network more easily repairable in the event of damage \cite{Albert2000, Farr2014}.



Central to all this work has been the notion of resilience or robustness \cite{Gao2016}. We refer to the resilience of a network as its ability to maintain functionality in the event of failures or environmental changes. Of course, this depends on how the functionality of a network is evaluated. One way of quantifying functionality is to track how network connectivity properties change as edges are gradually removed. Percolation theory has been used extensively to describe how different complex networks break down as they are subjected to increasing damage \cite{Cohen2000, Holme2002}. In a network that transports material, there is the additional consideration of how changes to the network structure alter the flow field. There has been increasing evidence suggesting that the flow of a transportation network influences the network structure \cite{Alim2018}. For instance, in the finest vessels of microvasculature, red blood cells cause transient occlusions that may be important in regulating uniform flow throughout the network \cite{Chang2017}. Moreover, networks are thought to evolve based on adaptive rules controlled by the edge flows that could be subject to transient occlusions \cite{Ronellenfitsch2016, Cornelissen2009}. In this work we explore the resilience of networks to single edge perturbations. If one network edge is occluded, flow will be rerouted around the occlusion resulting in a new equilibrium flow field. Depending on the edge capacities, the rerouting of flow may leave some edges overloaded, some under-supplied, and others with the flow direction reversed. All of these situations may be detrimental to network functionality. Moreover, depending on the network structure, the rerouting of flow can affect sites far from the perturbed link. For regular lattices, such as the square grid, the flow redistribution after a single edge is removed can be computed exactly by utilizing symmetries of the network \cite{Cserti2002}. For  disordered networks, a numerical approach is required. 

The problem of flow disruption after network damage has been heavily developed in the field of power engineering. When a line goes down in a power grid, the power that was once transferred by that line must be redistributed across the remaining lines. The altered flow can be computed using line outage distribution factors \cite{Wood2014}. Because these distribution factors are useful in predicting weaknesses in the network, considerable work has been done to optimize the calculation of distribution factors while minimizing the amount of resources used to monitor the system \cite{Guler2007, Ronellenfitsch2017a, Hines2017}. Line outage distribution factors have also been used to demonstrate the presence of Braess' paradox in power grids, where the addition of a single edge has the potential to cause overloads on distant edges \cite{Witthaut2013, Labavic2014}. The asymptotic form for line outage distribution factors has been derived for several network topologies, including small-world networks, in \cite{Bernasconi2017} and \cite{Strake2018}.
Here we focus on a class of networks with two distinct features: elements of disorder and a hierarchical structure of edge conductances. To our knowledge, the effects of damage in heterogeneous networks with high conductance veins have not yet been considered.

This work can also serve as a minimal model relevant to clots in brain vasculature. Brain vasculature forms a network with hierarchically ordered vessels: blood is routed from highly conductive surface cortical arteries to the intricate structure of micro-vessels that supply the brain tissue with oxygen via mid-sized penetrating arterioles \cite{Blinder2010}. The same equations that govern resistor networks have been used to describe vascular flow networks \cite{Lorthois2011a}. Several experimental studies have aimed to model the impact of flow redistribution on global brain functionality \cite{Nishimura2007, Lorthois2011}. Previous work on ischemic strokes has shown that the penetrating arterioles are especially vulnerable to damage because the network is unable to efficiently reroute flow after an obstruction \cite{Nishimura2010,Blinder2013}. The goal of our work in this context is to understand how highly conductive vessels, such as the penetrating arterioles that permeate the capillary bed of the cortex, change the redistribution of fluid flow when an occlusion forms close to the vessel. While real brain vasculature contains both veins and arteries, here we consider only half of the system, tracking the current from a single input point to distribution in the capillaries and venules, modeled as sinks. Colloquially we will refer to any highly conductive vessel as a vein, although it is understood that it can function as either a vein or artery.



\section{Calculation of Network Flow}

The basic calculation of flow redistribution after a local occlusion in a network is outlined here and presented fully in Appendix \ref{appendix:dip_calc}. Given a laminar, non-pulsatile flow network with edges $ij$ weighted by conductance $C_{ij}$ and with current $Q_i$ injected into node $i$, the goal is to determine the edge currents $I_{ij}$. This can be done exactly by solving for the vector of node potentials $\overline{v}$ using $\overline{\overline{L}}\overline{v} = \overline{Q}$ where $\overline{\overline{L}}$ is the graph Laplacian, then solving for $\overline{I}$ using $I_{ij} = C_{ij}(v_i-v_j)$. Since $\overline{\overline{L}}$ does not have full rank, the node potentials satisfying $\overline{\overline{L}}\overline{v} = \overline{Q}$ are not unique. A solution can be found by setting a reference node $v_k = 0$, which is equivalent to adding a uniform potential bias to all nodes, effectively setting $v_k$ to be ground. The remaining node potentials can be computed by inverting the truncated Laplacian matrix with the $k^{\text{th}}$ row and column removed \cite{Biggs1997}. Since $I_{ij}$ depends on the difference between two node potentials, the constant potential bias does not change the values of the edge currents. An alternate method is to calculate the Moore-Penrose pseudoinverse of the Laplacian and obtain $v = L^+ Q$.

\begin{figure}[h]
\begin{center}
 	\includegraphics[height=38mm]{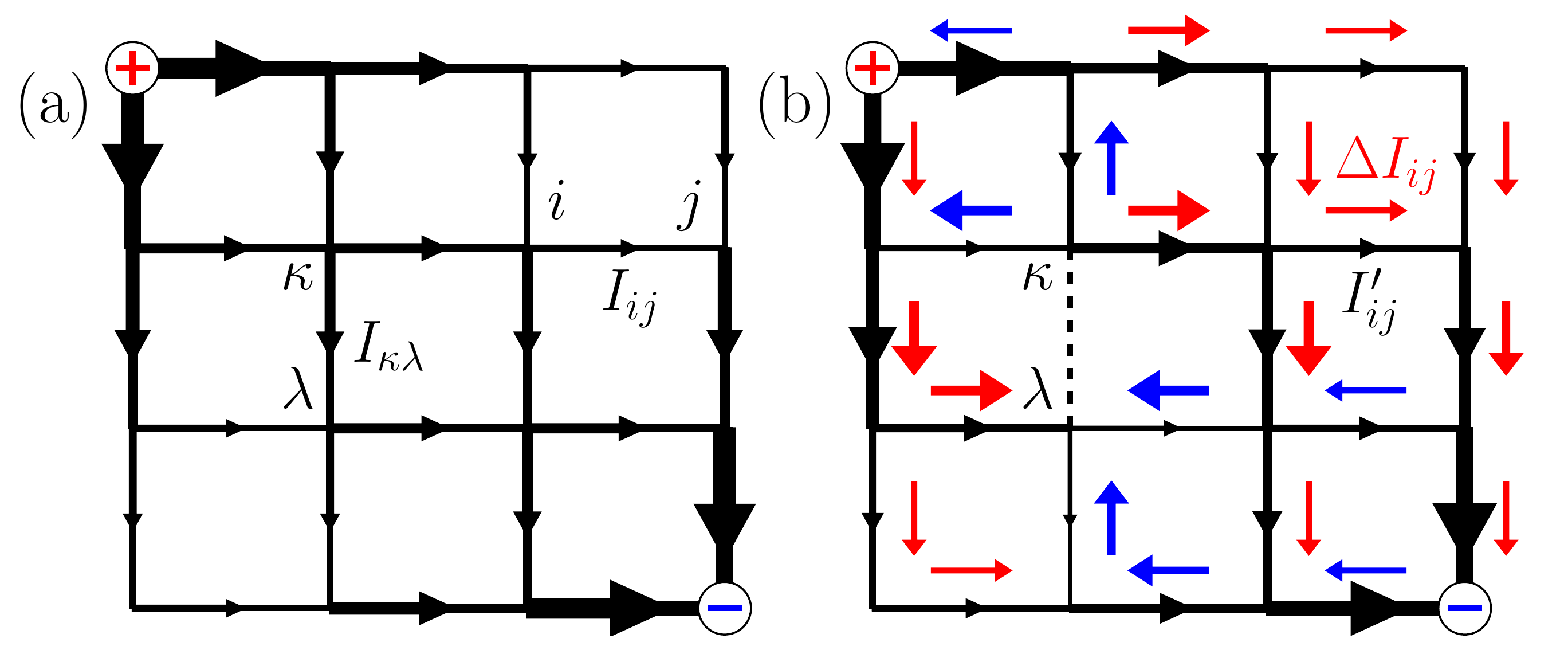}
    \includegraphics[height=42mm]{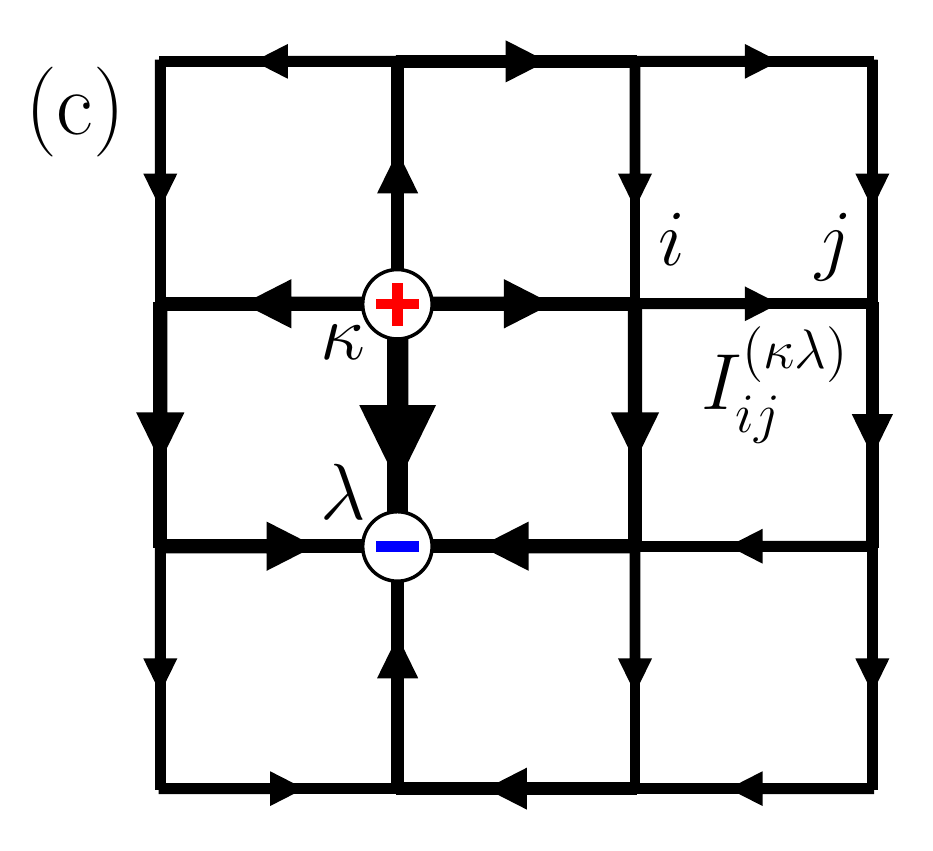}
\caption{A model flow network with one current source node (red cross) and one sink node (blue dash). (a)~The initial flow field $I_{ij}$. (b)~The new flow field $I'_{ij}$ after $C_{\kappa\lambda}$ is set to 0, with floating arrows illustrating the displaced current $\Delta I_{ij}$. Blue arrows indicate $I_{ij}\times\Delta I_{ij}<0$ ($I_{ij}$ and $\Delta I_{ij}$ are in the opposite direction) and red arrows indicate $I_{ij}\times\Delta I_{ij}>0$. ($I_{ij}$ and $\Delta I_{ij}$ are in the same direction). (c)~The undamaged network now with a dipole current source and sink, which we refer to as the $(\kappa\lambda)$ system. The black arrows in (c) are proportional to the colored arrows in (b), indicating that the displaced current field after damage is proportional to the current in the undamaged system with a dipole current source and sink.}
\label{cartoon}
\end{center}
\end{figure}

When the network is perturbed by blocking edge $\kappa\lambda$, that is, by setting $C_{\kappa\lambda} = 0$, the network edge currents change to a new flow field $I'_{ij}$. However, the final flow is not a sufficient measure to describe network flow disruption, since an occlusion may potentially reverse the direction of flow on an edge yet maintain the same flow magnitude. Because we define disruption as a significant change between the initial and final states of the edge flows, we instead track the displaced current, or the change in the current flow through each edge:
\begin{equation}
\Delta I_{ij} = I'_{ij} - I_{ij}
\end{equation}
Figure~\ref{cartoon}(a) shows an initial flow network with a single source and sink and Fig.~\ref{cartoon}(b) shows the final edge currents after the occlusion of edge $\kappa\lambda$ as well as the displaced current $\Delta I$ in floating arrows. $\Delta I_{ij}$ can either be positive, as seen on edges marked with red arrows, or negative, as seen on edges marked with blue arrows. Moving forward, we will use the absolute value of the displaced current as an indicator of disruption in network function.


While the difference between two flow fields is not necessarily intuitive, there is an alternate formulation that results in an equivalent form for $\Delta I$. We consider the original undamaged network, but now with a dipole current source and sink at nodes $\kappa$ and $\lambda$, which we refer to as the $(\kappa\lambda)$ system. Specifically, we set $Q^{(\kappa\lambda)}_\kappa = I_{\kappa\lambda}$ and $Q^{(\kappa\lambda)}_\lambda = -I_{\kappa\lambda}$, with zero injected current for all other nodes. The current flow $I^{(\kappa\lambda)}$ in this system is related to the differential current flow in the damaged system by
\begin{equation}
	\Delta I_{ij} = \frac{1}{1 - C_{\kappa\lambda}R_{\kappa\lambda}^{\text{eff}}} I_{ij}^{(\kappa\lambda)}.
\label{equiv_form}
\end{equation}
Here, $R_{\kappa\lambda}^{\text{eff}}$ is the effective resistance between nodes $\kappa$ and $\lambda$ in the original network and can be written as $R_{\kappa\lambda}^{\text{eff}} = L^{-1}_{\kappa\kappa} - 2L^{-1}_{\kappa\lambda} + L^{-1}_{\lambda\lambda}$ (see for instance \cite{Klein1993}). The full derivation of Eq.~\ref{equiv_form} is included in Appendix~\ref{appendix:dip_calc}, but this result is well-known. In the language of power engineering, this is the fact that the line outage distribution factor is proportional to the power transfer distribution factor \cite{Wood2014, Guler2007}. 

Equation \ref{equiv_form} can be observed by comparing the edge arrows in Fig.~\ref{cartoon}(c) to the floating arrows in Fig.~\ref{cartoon}(b) and noting that they qualitatively match. Moreover, tracing the edge arrows in Fig.~\ref{cartoon}(c) reveals a coarse-grained dipole pattern. In a large network far from the boundaries, the displaced current $\Delta I$ behaves like the electric field generated by a dipole charge, decaying as a power law as a function of distance from the damage site. Thus, far from the dipole current source (or equivalently, the damaged edge) $\Delta I \sim r^{-2}$ in a planar network and $\Delta I \sim r^{-3}$ in a 3D network. In this work we study exclusively 2D networks as a starting point to establish methodology. Examples of 2D or nearly 2D flow systems that experience perturbative damage can be found in leaf venation, slime molds, and retinal vasculature \cite{Snodderly1992}. The study of 3D networks has a wider range of biologically relevant applications and is reserved for future work.


\section{A Measure for Resilience}


Given the form of the displaced current $\Delta I$, here we present a way to quantify the extent of flow rerouting in order to compare the total network disruption for different damage sites. Our approach is to consider how far a test edge may be from the damaged edge and still experience a significant change in flow caused by the disturbance. We introduce the notion of \textit{edge tolerance}, defined as the normalized maximum displaced flow that an edge can sustain without being under-supplied, overloaded, or otherwise disrupted. We define the \textit{damage zone} for an edge to contain all edges that experience a change in flow exceeding their tolerance threshold. Specifically, upon inflicting damage to edge $\kappa \lambda$, the damage zone includes edges $ij$ that satisfy \begin{equation}
	\bigg| \frac{\Delta I_{ij}}{I_{\kappa\lambda}}\bigg| > t
\label{dz_def}
\end{equation}
where $t$ is a fixed threshold value. We normalize by the initial current flow at the damaged edge, which is the total amount of displaced flow that needs to be distributed among other network edges. With this in mind, $\Delta I_{ij}/I_{\kappa\lambda}$ is the damage sustained by edge $ij$ when edge $\kappa\lambda$ is occluded, and the edge is included in the damage zone if the damage sustained exceeds the damage tolerance. The left side of the inequality is referred to the linear outage distribution factor in the context of power grid engineering. Using Eq.~\ref{equiv_form} (see Appendix~\ref{appendix:dz_indep}), Eq.~\ref{dz_def} becomes
\begin{equation}
 	\bigg| \frac{ C_{ij} (L^{-1}_{i\kappa} - L^{-1}_{i\lambda} - L^{-1}_{j\kappa} + L^{-1}_{j\lambda})}
	{1 - C_{\kappa\lambda}(L^{-1}_{\kappa\kappa} - 2L^{-1}_{\kappa\lambda} + L^{-1}_{\lambda\lambda})} \bigg| > t.
\label{env_condition}
\end{equation}
Thus, the inclusion of edge $ij$ in the damage zone of edge $\kappa \lambda$ is dependent only on the threshold $t$, the conductances, and the connectivity of the graph (encoded in the Laplacian). In particular, the damage zone is independent of the initial current $I_{\kappa\lambda}$ and thus also the initial current sources and sinks, which is convenient because these are difficult to measure in a real biological flow network. Furthermore, because the damage zone is not sensitive to the net current, this metric truly probes the effects of network topology, or the properties that arise from the way the network is wired. The damage zone quantifies the extent of the network that is affected above the threshold by the damage of a single edge. A small damage zone signifies that the rerouted flow is contained within a small neighborhood around the damaged edge, whereas a large damage zone reflects that the rerouting is more spread out. A more resilient network minimizes the damaged region and thus is less affected by perturbative disturbances.

\begin{figure}[h]
\begin{tabular}{cc}
	\includegraphics[height=40mm]{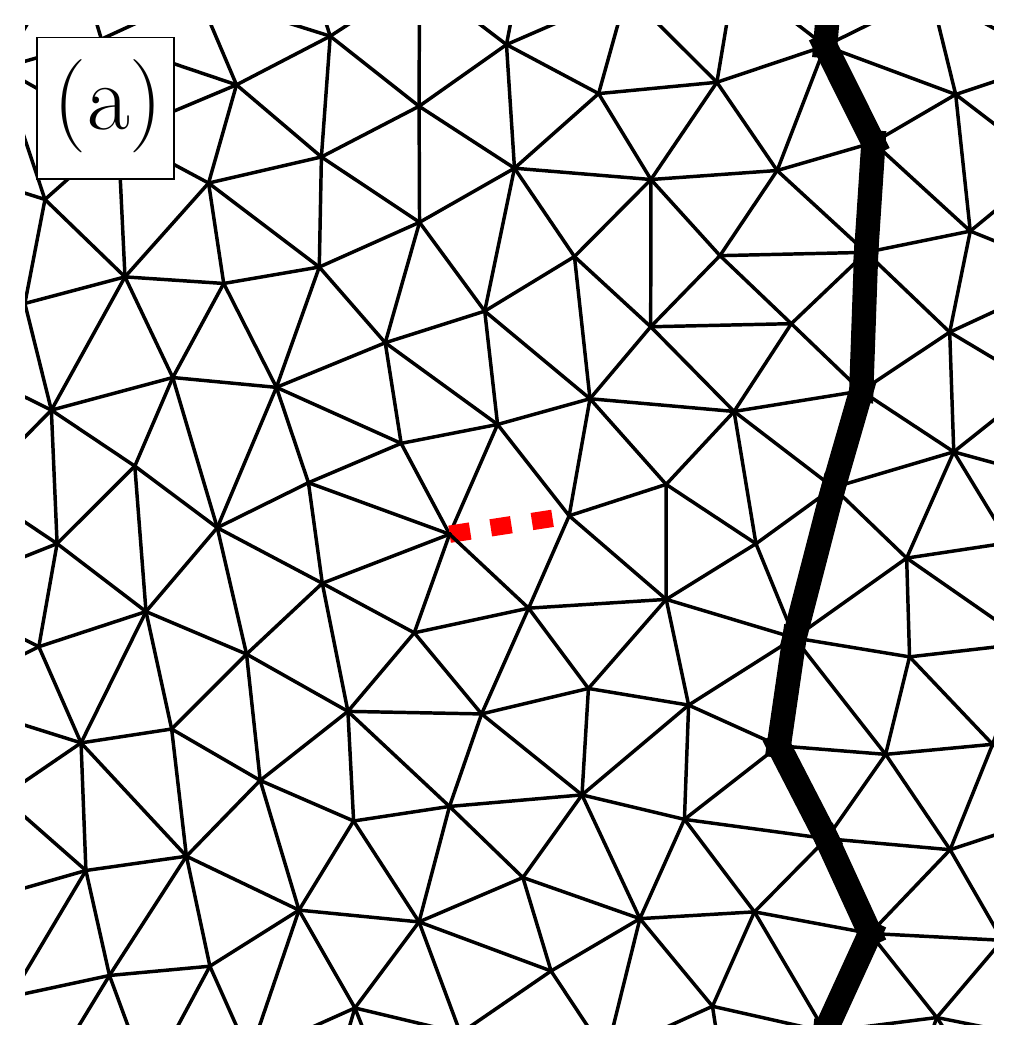} &
	\includegraphics[height=40mm]{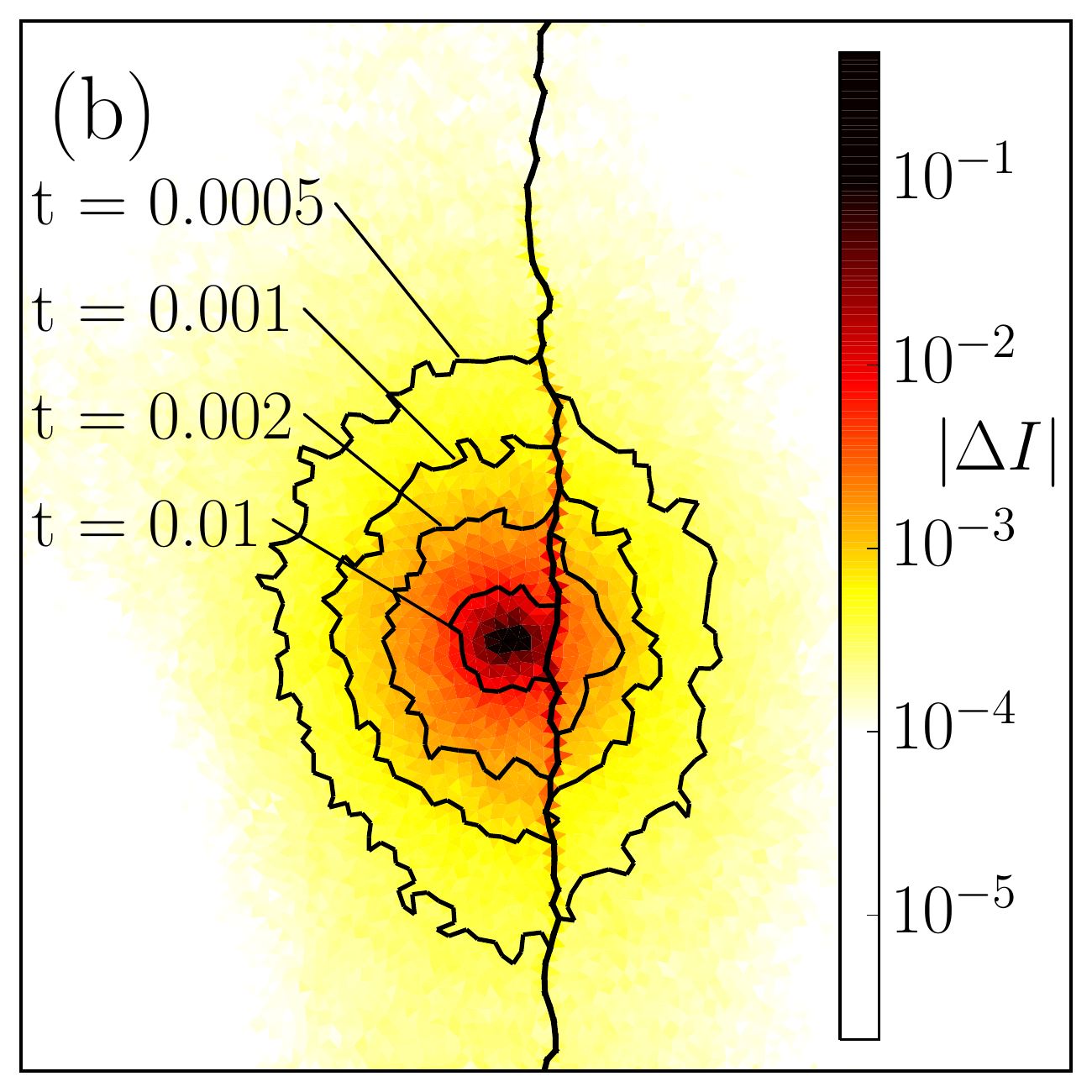} \\
\end{tabular}
\caption{An example of damage on a randomly packed (RP) triangular tiling with 5000 nodes. The bulk edges are set to conductance 1 and edges forming the central vein are set to conductance 5. (a) A close up view of the occluded edge (red dashed line) to show the network structure. (b) Contour lines indicate damage zones of different thresholds; edges within a boundary satisfy Eq.~\ref{dz_def} for different values of $t$. The shape of the damage zone becomes distorted upon crossing the vein.}
\label{singlesnip}
\end{figure}

This study intends to couple the local edge tolerance with the global network resilience. In this work we consider the resilience to be the ability of the network to withstand damage by minimizing the expected number of disrupted edges. In a realistic system, edges often have the ability to slightly change their conductance, for instance, by modulating the channel radius in response to a change in flow. This adaptive behavior complicates the problem of determining edge currents; this work will consider only systems with fixed edge conductances.

To study how a network feature such as a highly conductive vein changes the resilience, we examine the difference in the number of edges in the damage zone when a vein is present versus when it is absent. We first consider a simple structure: a single vein of increased conductance in a randomly packed (RP) triangular tiling network with otherwise uniform bulk conductance. The procedures for generating a RP triangular tiling and for drawing a highly conductive vein are outlined in Appendix~\ref{appendix:network}. Figure~\ref{singlesnip}(a) shows a typical network with a central vein and illustrates the shape of the damage zone for damage near the vein. Network sizes are chosen to be around 5000 nodes. 

We choose current boundary conditions that may be reasonable for a section of biological tissue: the node at the top of the vein is set to be a source of 1 unit of flow, the node at the bottom of the vein is set to be a sink for $1/2$ of the input flow, and all other nodes are set to be uniform sinks to accommodate the remaining $1/2$ of the input flow. This ensures that all edges have nonzero current flow and the network obeys net current conservation. Choosing different current sources and sinks does not significantly affect the current redistribution, as seen in the discussion of Eq.~\ref{env_condition}. Whereas the damage zone in a uniform network has a roughly circular shape, the damage zone near a vein is asymmetric, and the shape changes discontinuously upon crossing the vein. The vein serves to decrease the damage zone on the unperturbed side of the network, providing a shielding effect. The damage zones for four different threshold values are shown. For all following work, we fix $t = 0.005$, so the typical area covered by a damage zone is $\sim 1 \%$ of the total network area. Results qualitatively apply to a range of $t$ values, as shown in Appendix~\ref{appendix:thresh}.

If the occluded edge is sufficiently far from a vein then the damage zone will be the same regardless of whether or not the vein is present. However, if the edge is close by, the vein in the network will alter the damage zone. We calculate the \textit{edge shielding} $s_{ij}$ by counting the number of edges $N_{ij}$ in the damage zone for a removed edge $ij$ for a network with the vein present and again for a network with the vein absent, then taking the normalized difference:
\begin{equation}
s_{ij} = \frac{N_{ij, \text{vein present}}}{N_{ij, \text{vein absent}}} - 1
\end{equation}
If $s_{ij} < 0$ the presence of the vein has decreased the damage zone, increasing the global network resilience. 

\section{Results}

\subsection{Veins Provide Shielding}
\label{one_vein}

\begin{figure}
\includegraphics[width=84mm]{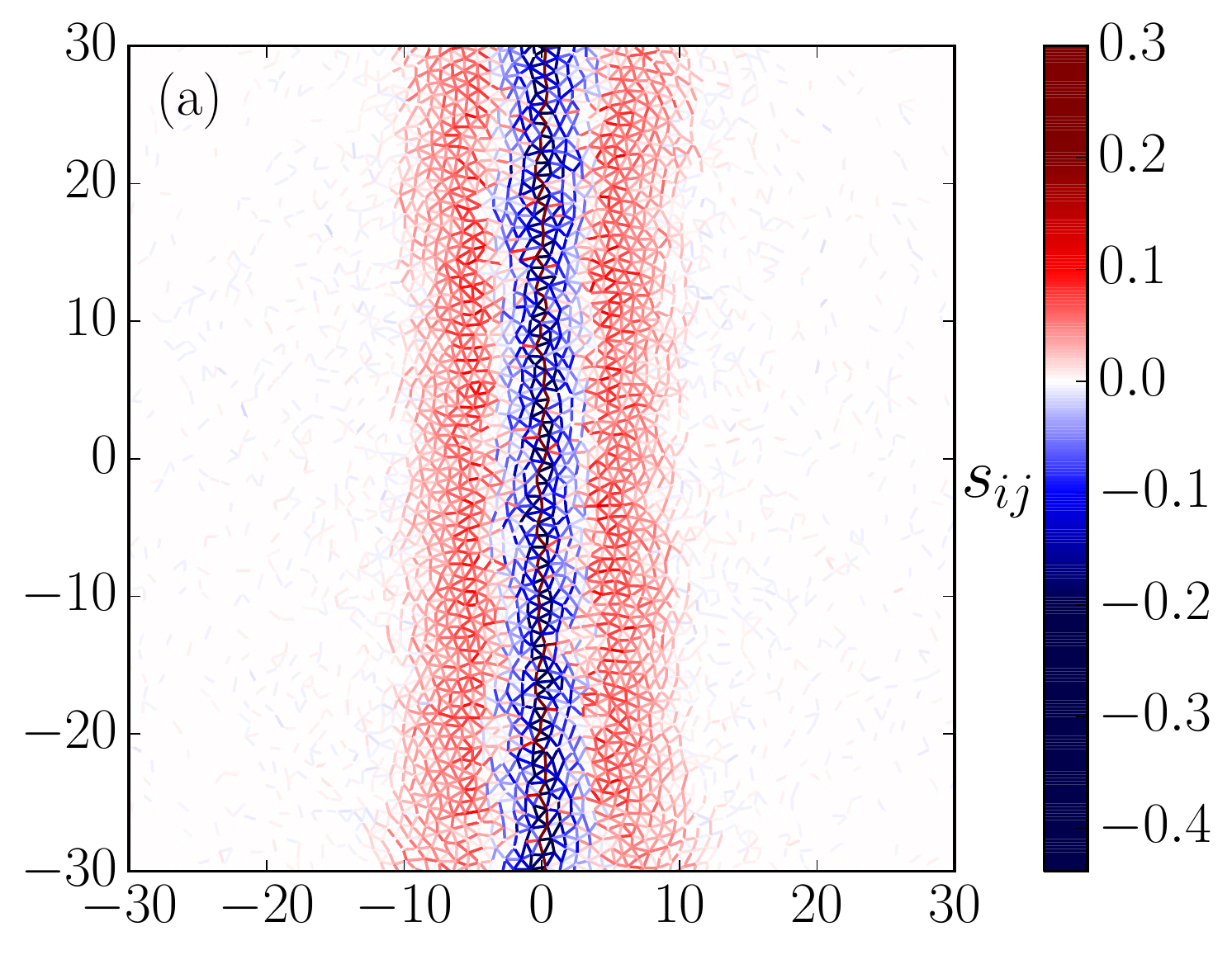}
\label{fig:label:a}
\hspace*{-1.9cm}\includegraphics[width=72mm]{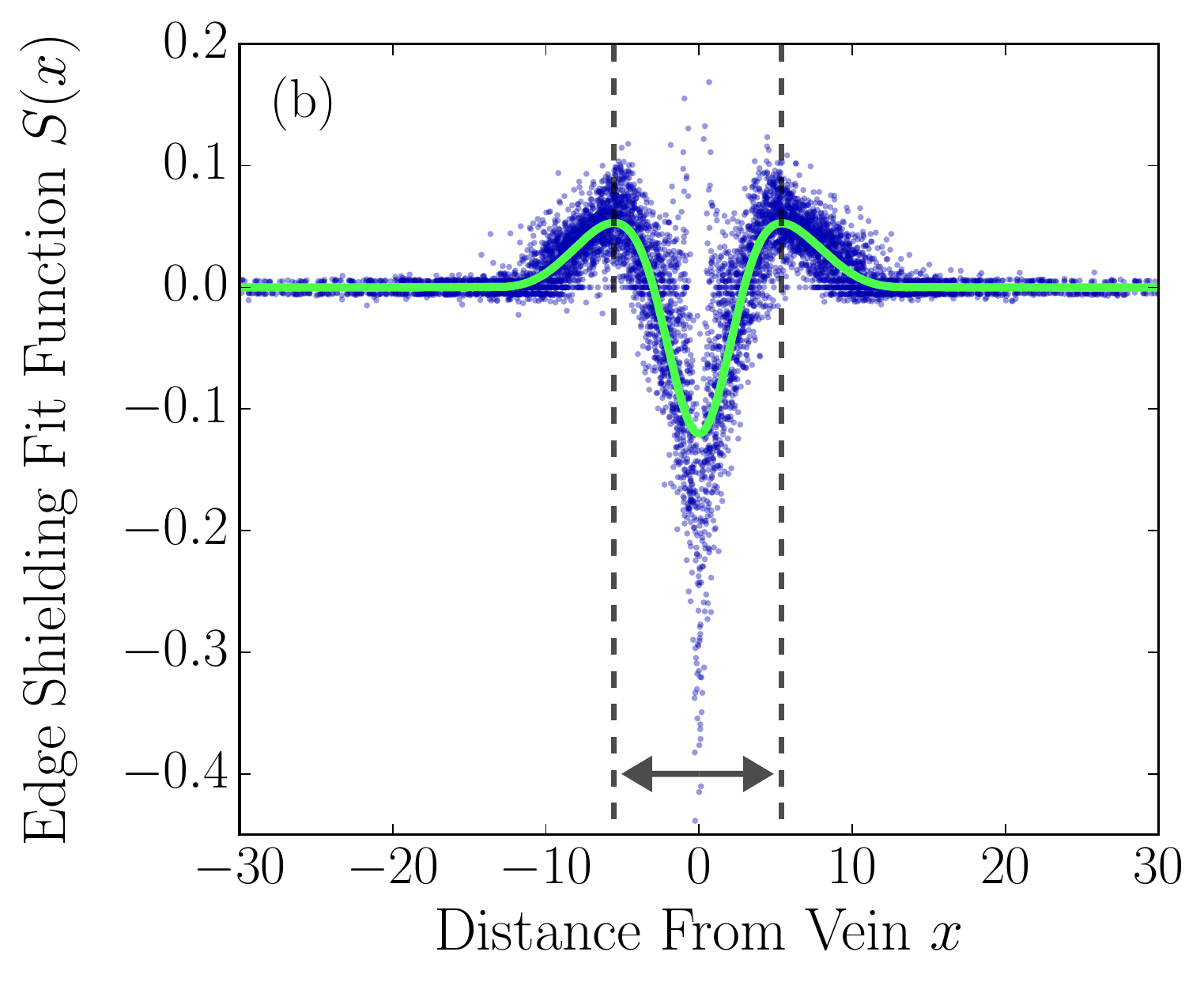}
\label{fig:label:b}
\caption{A RP triangular tiling with 5500 nodes, edges of conductance 1, and a single vertical vein through the center of the network composed of edges with conductance 5. The effect of the vein on the global network resilience is shown by computing $s_{ij}$ for each edge individually. (a) The real-space network with edges colored by $s_{ij}$. The vein provides shielding for edges with $s_{ij} < 0$ (blue) but increases the damage zone for edges with $s_{ij} > 0$ (red). Distances are measured from the network center in units of mean edge length. (b) The edge shielding fit function $S(x)$ is attained by projecting $s_{ij}$ onto the x-coordinate of the edge center and fitting the binned data using GPR. Edges on the vein are omitted from the plot and the calculation of $S(x)$. The shielding length $L_s$ is defined as half of the distance between two maximum of the $S(x)$.}
\label{shield}
\end{figure}

We study how adding a highly conductive vein affects the global network resilience by computing $s_{ij}$ for each edge individually and analyzing the distribution of $s_{ij}$ across the full network. Fig.~\ref{shield}(a) shows $s_{ij}$ in the real space of the network and Fig.~\ref{shield}(b) shows what we will refer to as the edge shielding fit function $S(x)$, which is a fit to the x-coordinate projection of $s_{ij}$ for each edge. Spatial dimensions are expressed as the Euclidean distance in units of the mean edge length. The center of the network lies on the origin and the vein lies roughly on the line $x = 0$. In this convention, the x-position of an edge can be positive or negative, depending if it lies to the right or to left of the vein respectively. 
The distribution of $s_{ij}$ forms alternating regions of negative and positive value around the vein. Edges on the vein will always have $s_{ij} > 0$. Close to the vein edges tend to have $s_{ij} < 0$, meaning that the vein decreases the size of the damage zone for these edges. Beyond this, there are two significant regions of the network where $s_{ij} > 0$ and the vein increases the damage zone. Far from the vein $s_{ij} = 0$, and the effects of the vein decay.

We define the \textit{shielding length} $L_s$ for a vein as the distance from the vein at which $S(x)$ achieves a maximum, beyond which it gradually decays to zero. More precisely, the shielding length is defined as half of the distance between the two maxima on either side of the vein. Qualitatively, the shielding length is the distance that the vein shielding effects extend on the network: an edge with $-L_s < x < L_s$ will likely experience a reduced damage zone when the vein is present, but an edge beyond this distance can only experience an increased damage zone. Another candidate for $L_s$ is the zero crossing of $S(x)$, but we use the location of the peaks because firstly, the maxima are strong features that can be identified independently of the type of fitting procedure used and secondly, because the positive region of the curve is a prominent feature of $S(x)$ indicating a strong vein effect and should not be excluded. The fit for $S(x)$ is produced by projecting $s_{ij}$ onto the x-axis, and binning these values using a bin size set to the mean network edge length. We model the binned data using Gaussian process regression (GPR). GPR succeeds in fitting to the two maxima, whereas a simple spline fails to provide reliable fits primarily due to the sharp minimum inherent to $S(x)$. Generally GPR fits have a coefficient of determination greater than 0.90, and typically poorer fits are due to lattice effects in the more symmetric networks. On-vein edges pose a problem for calculating the shielding signal since they are highly positive and thus interfere with the negative regions of $S(x)$. Since we are primarily interested in the behavior of the edges surrounding the vein, the on-vein edges are excluded from the fitting. 

\begin{figure}
\begin{center}
\begin{tabular}{cc}
	\includegraphics[height=45mm]{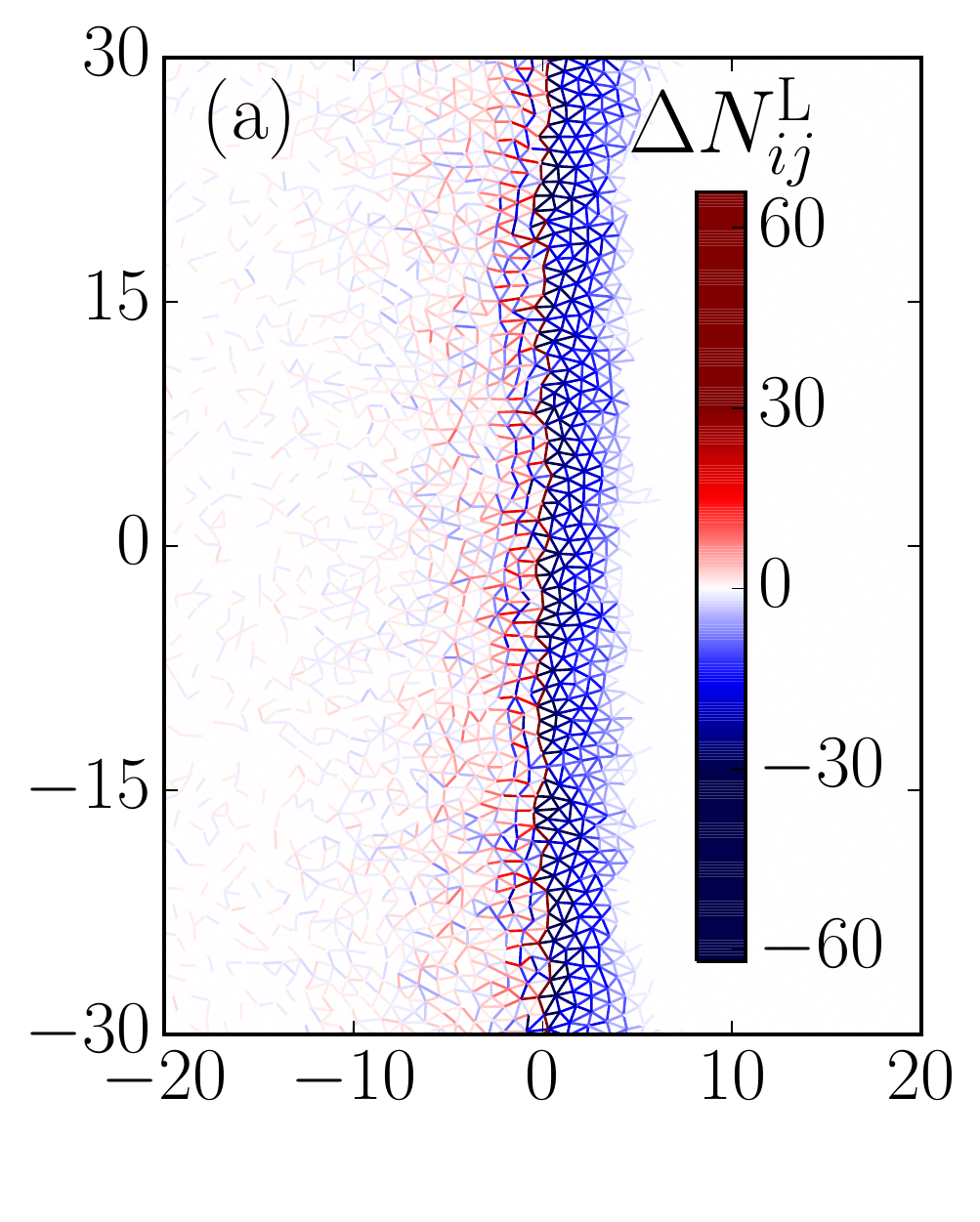} &
	\includegraphics[height=45mm]{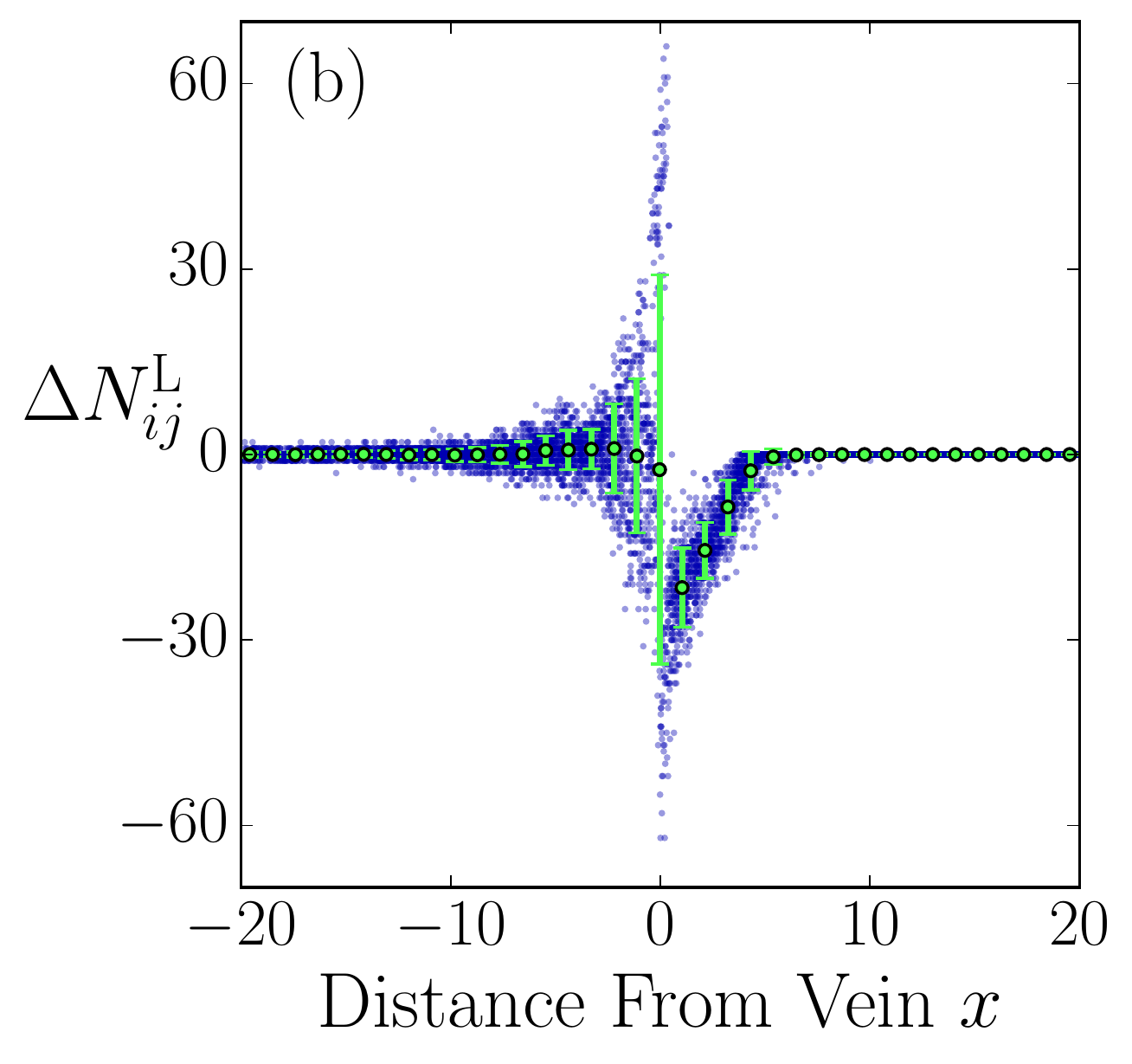} \\
\end{tabular}
\caption{Difference of damage zone edges to the left of the vein, $\Delta N_{ij}^{\text{L}}$, shown in (a) the real-space network and (b) the x-coordinate projection, with green points showing the binned averages, with error bars indicating one standard deviation of the bin. Edges with $0 < x < 5$ have $\Delta N_{ij}^{\text{L}} < 0$, indicating the vein serves as a conduit to absorb displaced flow, preventing it from leaking to the opposite side. This negative region explains the minimum in Fig.~\ref{shield}.}
\label{leftRight}
\end{center}
\end{figure}

We use the damage zone as a method of inferring how the flow in the system changes when a vein is added to the network. Edges in the damage zone have experienced the greatest amount of flow change. Without focusing on the details of how the current is rerouted, looking at changes in the damage zone will explain how the vein affects the flow. We have identified edges for which adding a vein results in a significantly changed damage zone, but now we want to see where this change comes from. We will examine which areas of the damage zone contribute most to the overall change, and why the damage zone is increased for some edges and decreased for others. This will reveal the mechanism behind edge shielding.

\begin{figure}[h]
\begin{center}
\begin{tabular}{cc}
\includegraphics[height=45mm]{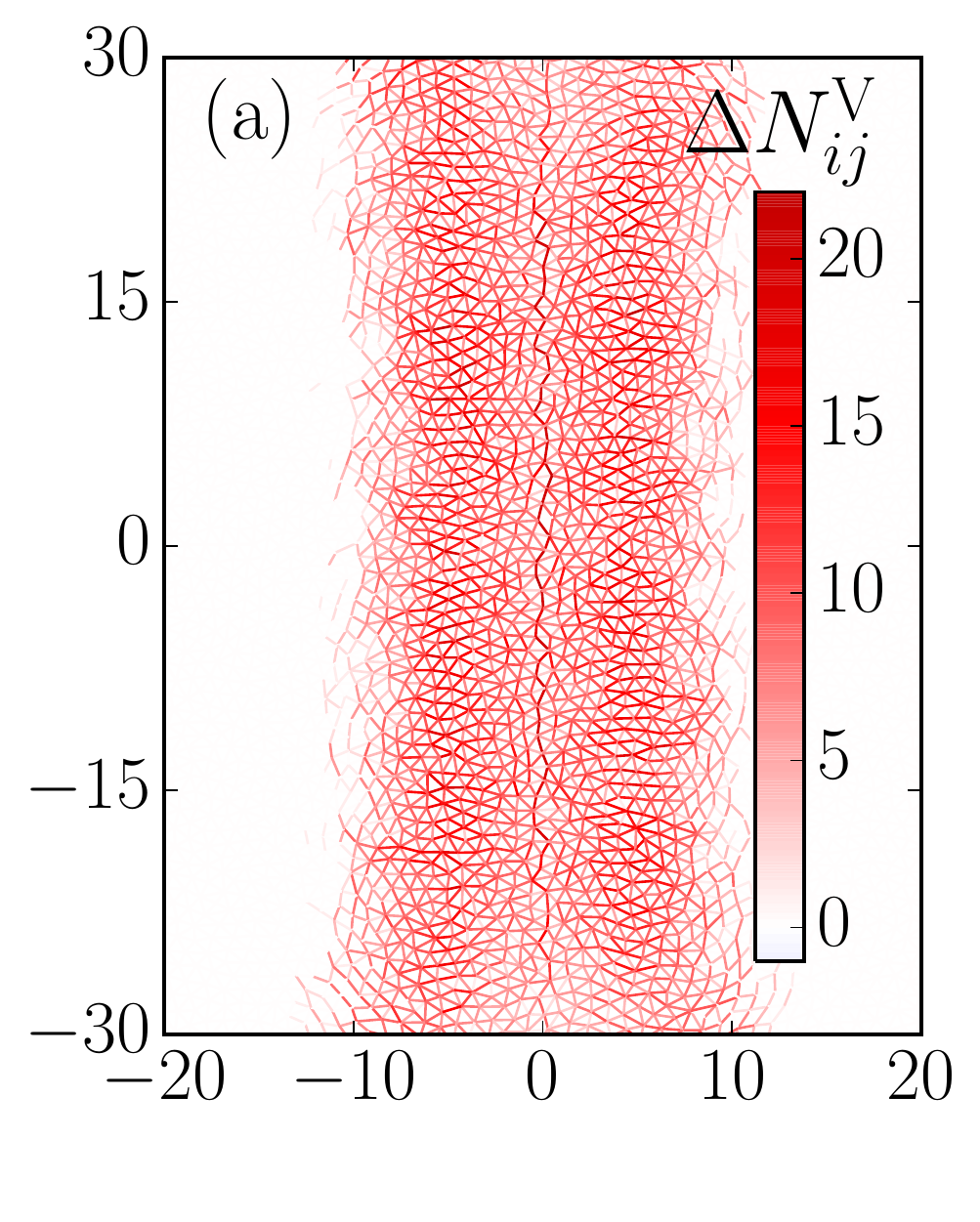} &
\includegraphics[height=45mm]{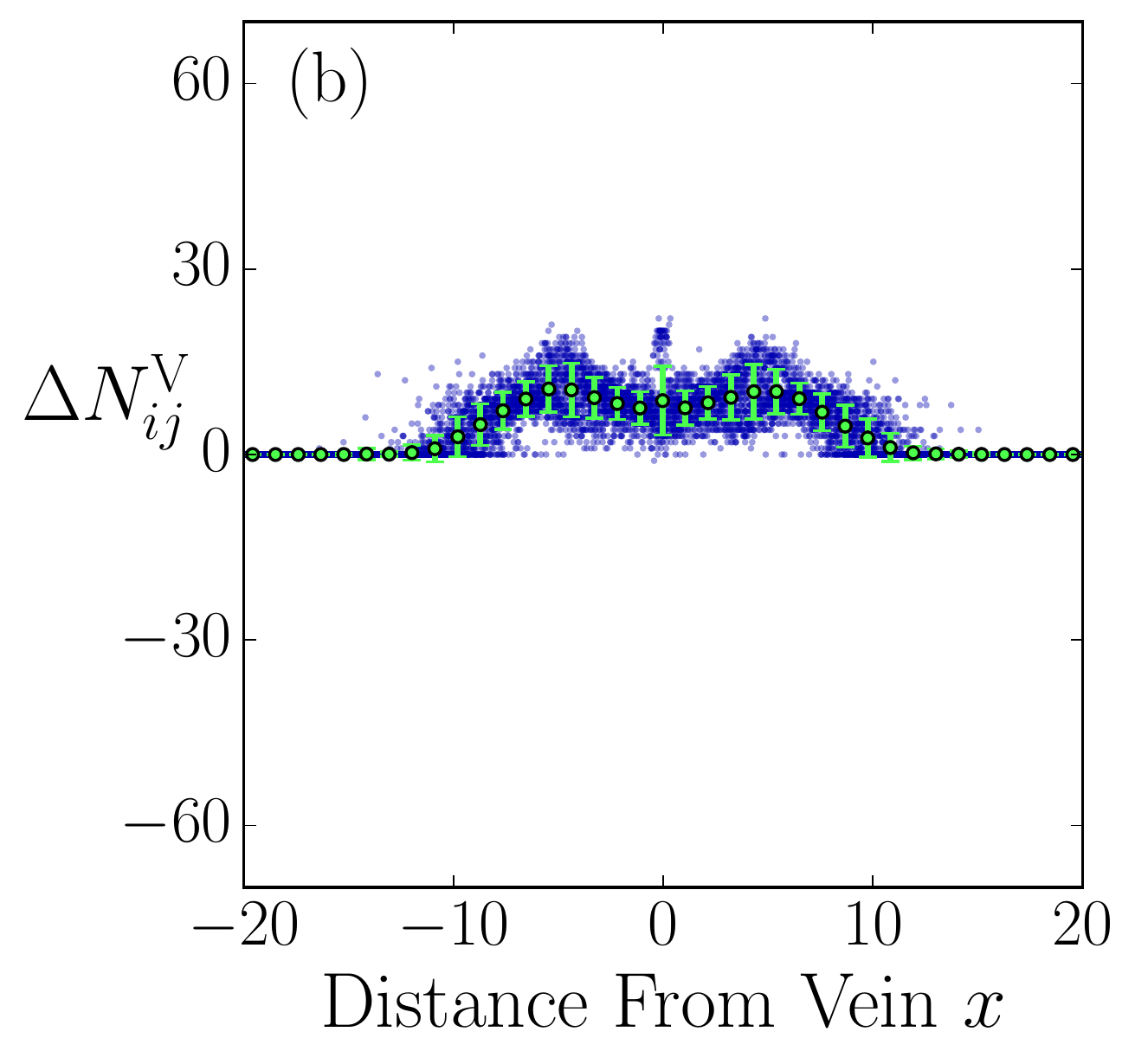} \\
\end{tabular}
\caption{Difference of damage zone on the main vein, $\Delta N_{ij}^{\text{V}}$, shown in (a) the real-space network and (b) the x-coordinate projection, with green points showing the binned averages, with error bars indicating one standard deviation of the bin. Axis scales and color bars are as in Fig.~\ref{leftRight} for comparison. The effects of the vein hold for $-11 < x < 11$, which is a greater range compared to Fig.~\ref{leftRight}, although the magnitude is smaller. This effect explains the two maxima of $S(x)$ in Fig.~\ref{shield}.}
\label{zero}
\end{center}
\end{figure}

We separate the damage zone into three populations of edges: $N_e = N_{ij}^{\text{L}} + N_{ij}^{\text{V}} + N_{ij}^{\text{R}}$, where $N_{ij}^{\text{L}}$ is the number of edges to the left of the vein, $N_{ij}^{\text{V}}$ is the number of edges on the vein, $N_{ij}^{\text{R}}$ is the number of edges to the right of the vein. For a veinless network, we draw an imaginary boundary where the vein would have been, so the three populations are still well-defined. In Fig.~\ref{leftRight} we consider only edges that lie on the left side of the damage zone; an analogous plot may be drawn for the right side. To avoid problems with division by zero, we plot $\Delta N_{ij}^{\text{L}} = N_{ij, \text{vein present}} ^{\text{L}}- N_{ij, \text{vein absent}}^{\text{L}}$, the unscaled difference in number of edges on the left side of the damage zone, as opposed to $s_{ij}$, the percentage difference in the number of edges. All edges on the right side of the vein have $\Delta N_{ij}^{\text{L}} \leq 0$. This means that crossing the vein significantly shrinks the damage zone, effectively shielding the damage. This shielding effect in which $\Delta N_{ij}^{\text{L}} < 0$ holds for edges with $0 < x \lesssim 5$. Edges on the left side of the vein attain both positive and negative values of $\Delta N_{ij}^{\text{L}}$. This means that if there is damage on the left side of the network, the left side of the damage zone may increase or decrease while the right side of the damage zone always shrinks. However, the mean amplitude on the left side of the vein remains close to zero, as seen by the green averaged points. 

The most prominent effect of the vein is to prevent displaced flow due to damage on the right side of the vein from crossing over to the left side, and vice versa. This is the cause of the observed edge shielding, and it is visualized in Fig.~\ref{shield}(b) by the deep minimum of $S(x)$ centered at the vein. The vein provides a low resistance channel to reroute displaced current for nearby damage, containing flow in the vein edges and preventing it from leaking to the other side of the vein. The high conductance of the vein allows the network to use a smaller portion of edges to reroute flow. 

\begin{figure}[h!]
\begin{center}
\begin{tabular}{cc}
\includegraphics[height=50mm]{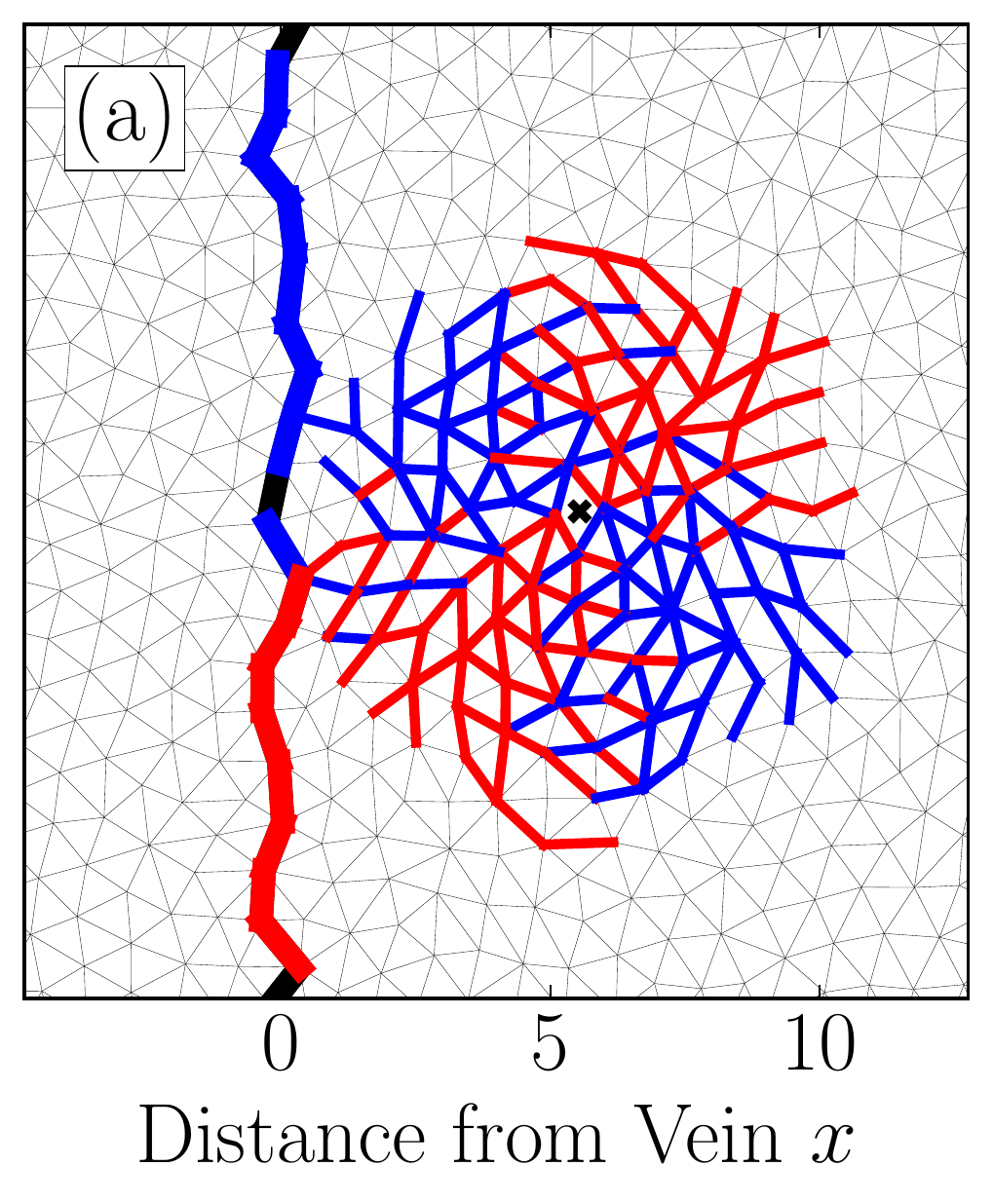} &
\includegraphics[height=50mm]{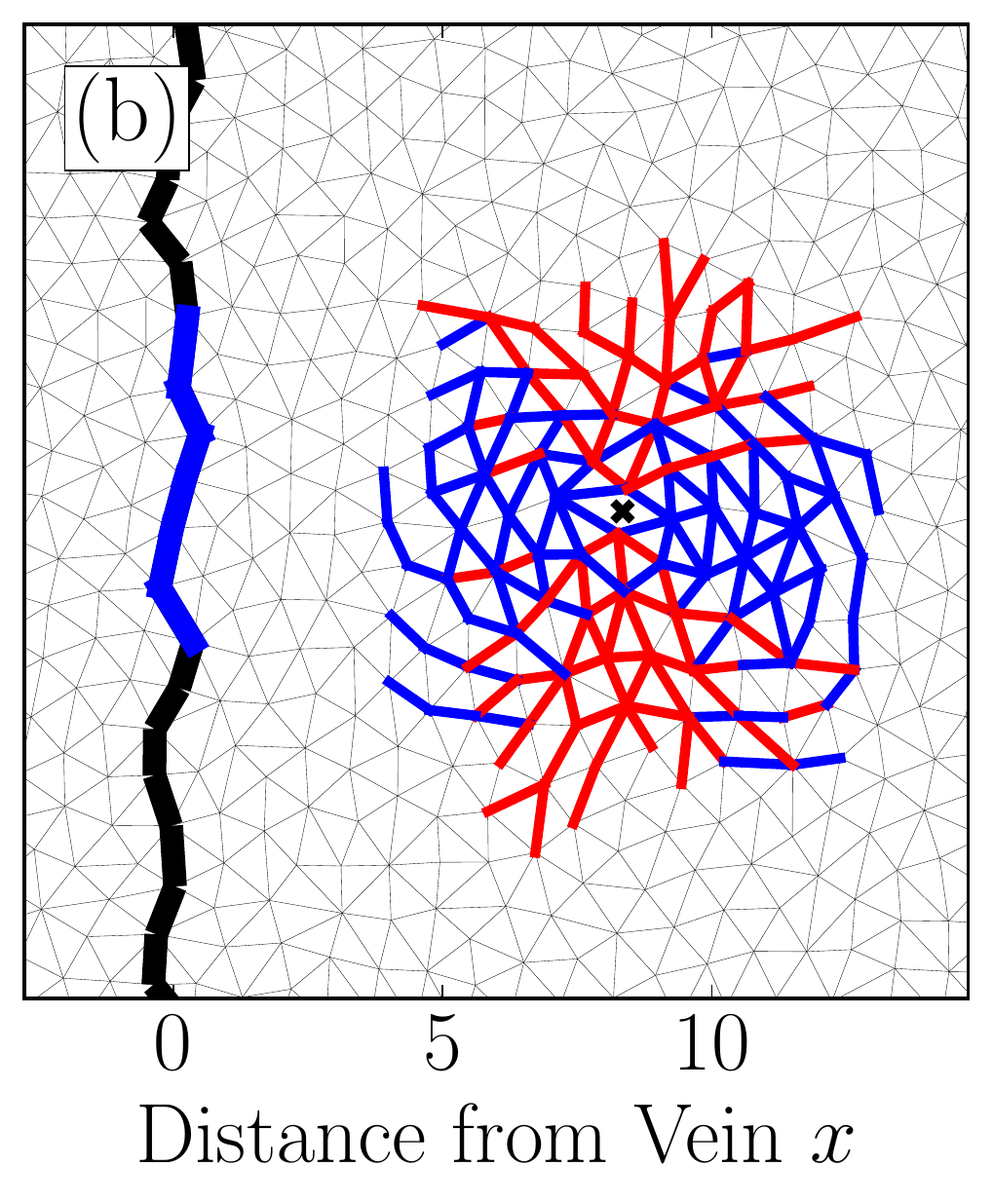} \\
\end{tabular}
\caption{The damage zones for two damage sites (denoted with a small cross), one close to the vein (a) and one far from the vein (b). As in Fig.~\ref{cartoon}, edges with $I_{ij} \times \Delta I_{ij} < 0$ are colored blue and edges with $I_{ij} \times \Delta I_{ij} > 0$ are colored red. The damage zone may be divided into two distinct parts: edges on the vein and edges enveloping the damage site.}
\label{DZ_examples}
\end{center}
\end{figure}

However, adding a vein does not decrease the damage zone for all network edges. The positive regions in Fig.~\ref{shield} are edges for which the damage zone has been increased. To explain these regions, we once again split the damage zone into three populations of edges, now plotting $N_{ij}^{\text{V}}$, the number of edges that lie on the vein, in Fig.~\ref{zero}. The main observation is that edges with $-11 < x < 11$ will have on-vein edges in their damage zones. Compared with the effect seen in Fig.~\ref{leftRight} which only persists for $x<5$, this is a long range effect. Even for distant damage, the vein actively serves to reroute flow. Individual inspection of the full damage zone for two sample edges is shown in Fig.~\ref{DZ_examples}. The on-vein edges of the damage zone are a distinct component, clearly separated from the part of the damage zone that envelopes the damage site. 

This discontinuity can be explained by Eq.~\ref{env_condition}, the damage zone threshold condition written in terms of edge flows and conductances. The denominator is independent of $ij$ and thus constant for all edges. The term $L^{-1}_{i\kappa} - L^{-1}_{i\lambda} - L^{-1}_{j\kappa} + L^{-1}_{j\lambda}$ is dependent primarily on distance from the damage site, as information about $C_{ij}$ is lost in the matrix inversion. Thus, this term does not distinguish the highly conductive vein edges. However, the other term in the numerator, $C_{ij}$, is of course sensitive to the vein and is able to bump on-vein edges beyond the damage zone threshold even if they are at a further distance. In other words, Eq.~\ref{dz_def} is likely to be satisfied because $\Delta I_{ij}$ is large compared to $I_{\kappa\lambda}$, even though $\Delta I_{ij}$ is small compared to $I_{ij}$. This means that our model is likely to qualify an on-vein edge $ij$ as a significant disturbance in the system when in fact the percentage change in current through $ij$ is quite small. Effectively, in this model, the tolerance of an on-vein edge and an off-vein edge have been set to equal values. A more realistic model should possibly scale edge tolerance with conductance. For the current model, the low tolerance of the vein edges is the reason for large regions of $S(x)>0$.

One final note is that superimposing Fig.~\ref{leftRight} with its mirror image and with Fig.~\ref{zero}, effectively summing all three parts of the damage zone that we had previously separated, recovers Fig.~\ref{shield}, up to normalization. This decoupling of the damage zone is essential for explaining the short-range negative region, the mid-range rise, and the far range decay of the edge shielding. Although we do not derive a functional form for $S(x)$ we can explain each feature separately through the behavior of the damage zone at different distances from the vein.

\subsection{Shielding is Controlled by Topology}
\label{shielding}

\begin{figure}
\begin{center}
	\includegraphics[width=84mm]{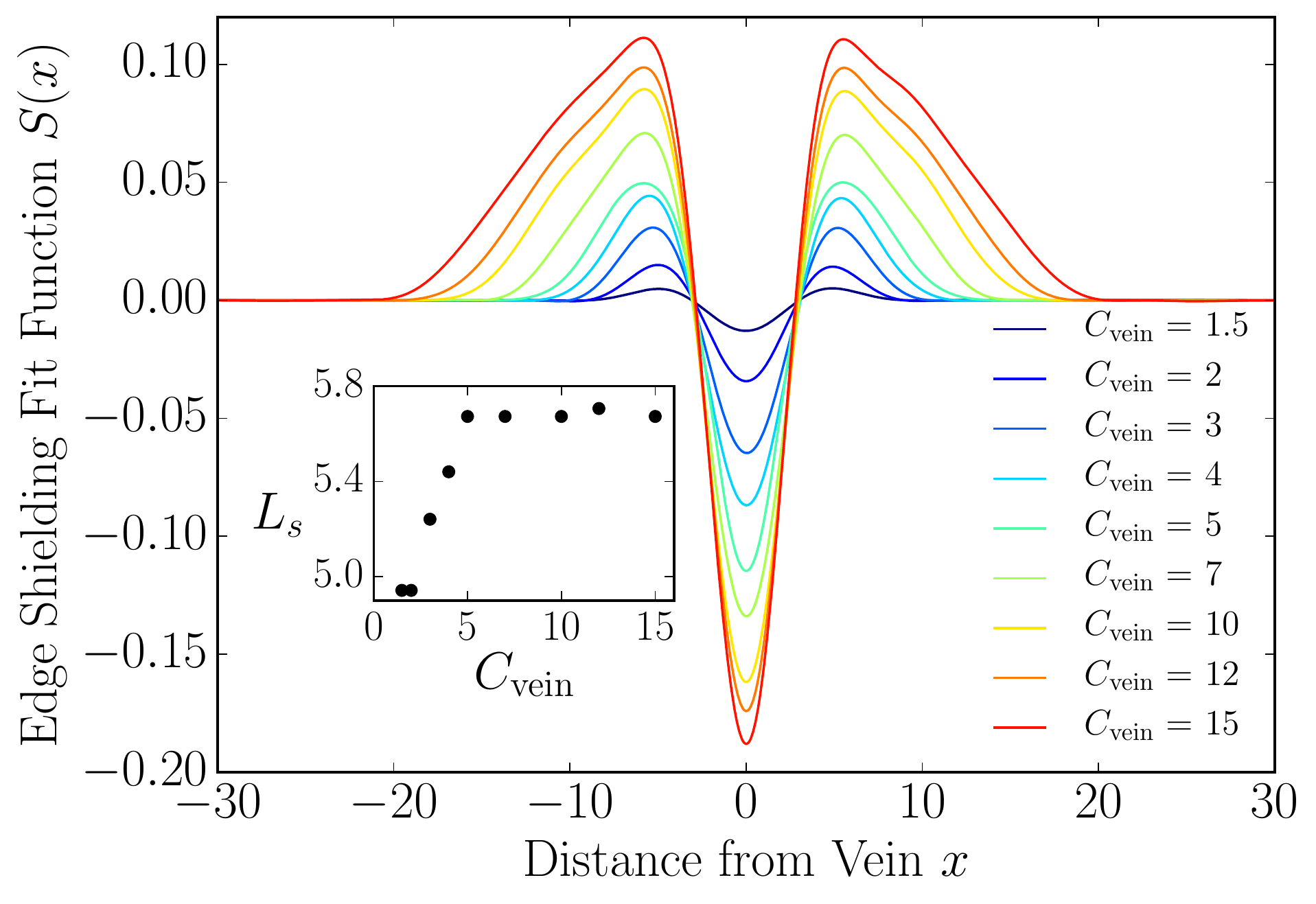}
\caption{Edge shielding fit functions $S(x)$ for a RP triangular tiling with 5500 nodes with bulk edge conductance 1 and a central vein set to 10 different conductance values $C_{\text{vein}}$. Inset: the shielding length $L_S$ appears clearly by $C_{\text{vein}} = 2$, increases until $C_{\text{vein}} = 5$, then becomes asymptotic. For $C_{\text{vein}} = 1.2$, $S(x)$ is nearly flat, so the maxima are not clearly distinguished and the derived $L_s$ is unreliable.} 
\label{growVein}
\end{center}
\end{figure}

\begin{figure*}
\begin{center}
\includegraphics[width=180mm]{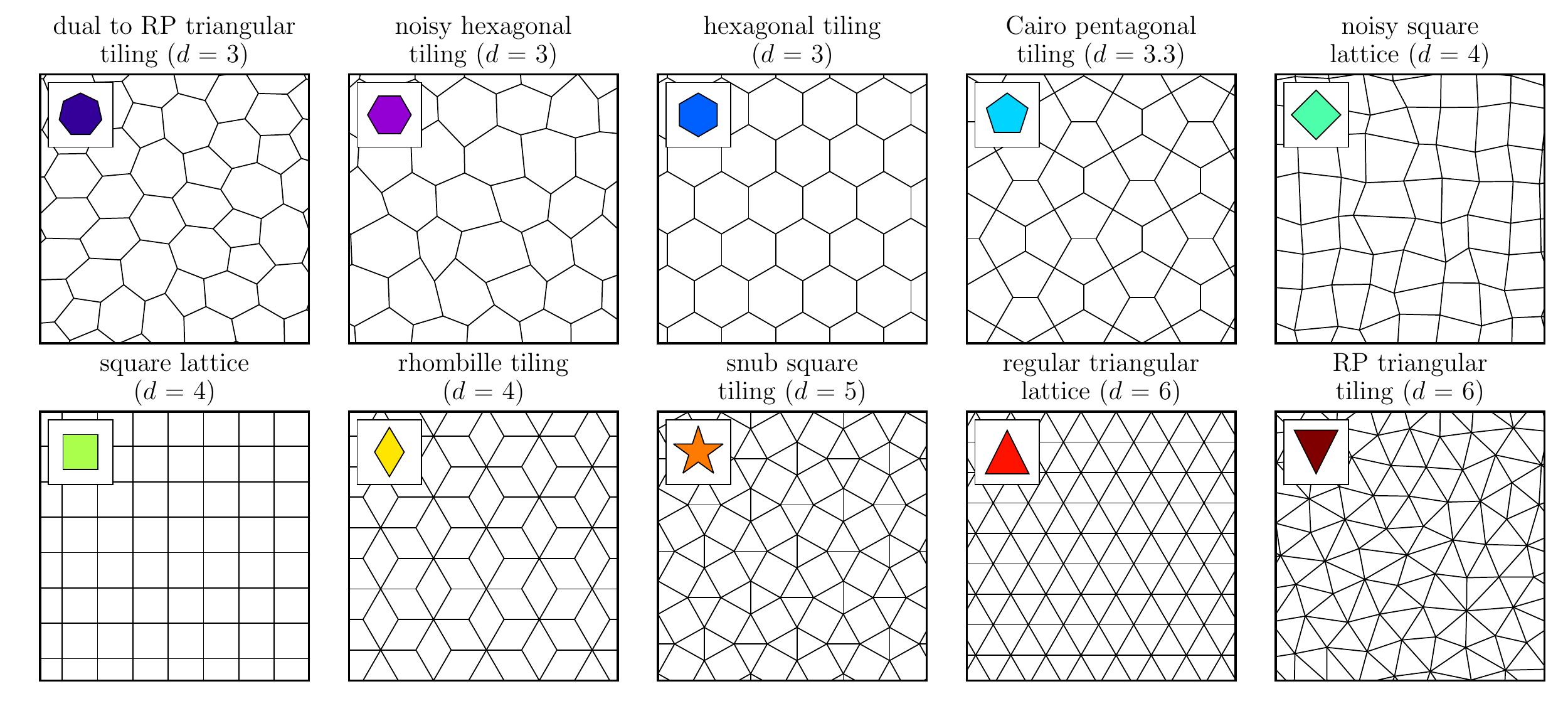}
\caption{Segments of the lattices used in Fig.~\ref{SL_topology}. The mean network degree $d$ varies between 3 and 6. Both regular lattices and networks with elements of disorder and noise are used. The edge conductance is set to the edge length scaled by the mean edge length of the network, so the average edge conductance is 1.}
\label{lattices}
\end{center}
\end{figure*}

In this section we ask what network properties control the shielding effects of the vein. We show how the shielding length is dominated by network topology, as opposed to geometry. An example of changing the geometry of a network includes changing edge conductances while preserving their relative hierarchy, or in other words, not suddenly making a bulk edge thicker than a vein edge. Changing the topology of the network entails a more severe modification to the underlying network connectivity, such as removing edges or growing additional veins. We will provide two examples of changes to the network geometry (thickening the central vein and increasing the network size) that do not significantly impact the shielding length. Then we will show that the shielding length is governed by a topological property of the network, namely the average degree of the nodes.

The first surprising result is that the shielding length is not controlled by the vein conductance. The edge shielding fit functions for networks with one central vein of increasing conductance $C_{\text{vein}}$ are shown in Fig.~\ref{growVein} and the inset tracks the shielding length $L_s$. Fits are obtained as in Fig.~\ref{shield}b and $L_s$ is defined as half of the distance between the two maxima. The shielding effect of the vein is characterized by a central minimum and two maxima. This effect is first noticeable once the vein conductance reaches 1.5 times the value of the bulk network conductance. However, after a relatively short period of growth, $L_s$ asymptotes to the constant value $L_s = 5.6$. This means that the positive shielding effects yield diminishing returns. When the vein conductance is four times larger than the bulk conductance, only edges with $-5.5 < x < 5.5$ experience a significant edge shielding $s_{ij}$ and increasing the vein conductance further does not significantly increase the shielding length. While the location of the maxima and the zeros of the shielding stay constant, the magnitude of $S(x)$ grows with increasing conductance. This means that the magnitude of shielding felt by edges within $L_s$ increases, and also that the long-range effects of the vein persist over a longer scale. For a fixed lattice, the edges that feel a shielding effect can be predicted by their  distance to the damaged vein, making the shielding length a topological effect.

As a second probe of shielding effects, we study a central vein in networks of varying size and topology. Previously we have just considered a single type of network: the RP triangular tiling. Now we extend our arsenal to include nine additional types of networks, shown in Fig.~\ref{lattices}. We are particularly interested in networks with a mean node degree between 3 and 4, which is typical for biological networks~\cite{Durand2006}. For each network we set the edge conductance to the edge length scaled by the mean edge length of the network, so the average edge conductance is 1. The standard deviation of edge conductances is $0.2$ for the noisy square lattice and around $0.15$ for other nonuniform lattices. Setting the edge conductance proportional to edge length is used to distinguish between the square lattice and the noisy square lattice, but the variation in edge conductances is small enough that results for a network with the variable edge conductances are similar to a network with uniform edge conductances. A central vein of conductance 5 is drawn for each network following the procedure in Appendix \ref{appendix:network}. For each network type we compute the shielding length $L_s$ for networks of increasing size, ranging from 10,000 to 40,000 edges. 

\begin{figure}[h!]
\begin{center}
	\includegraphics[height=55mm]{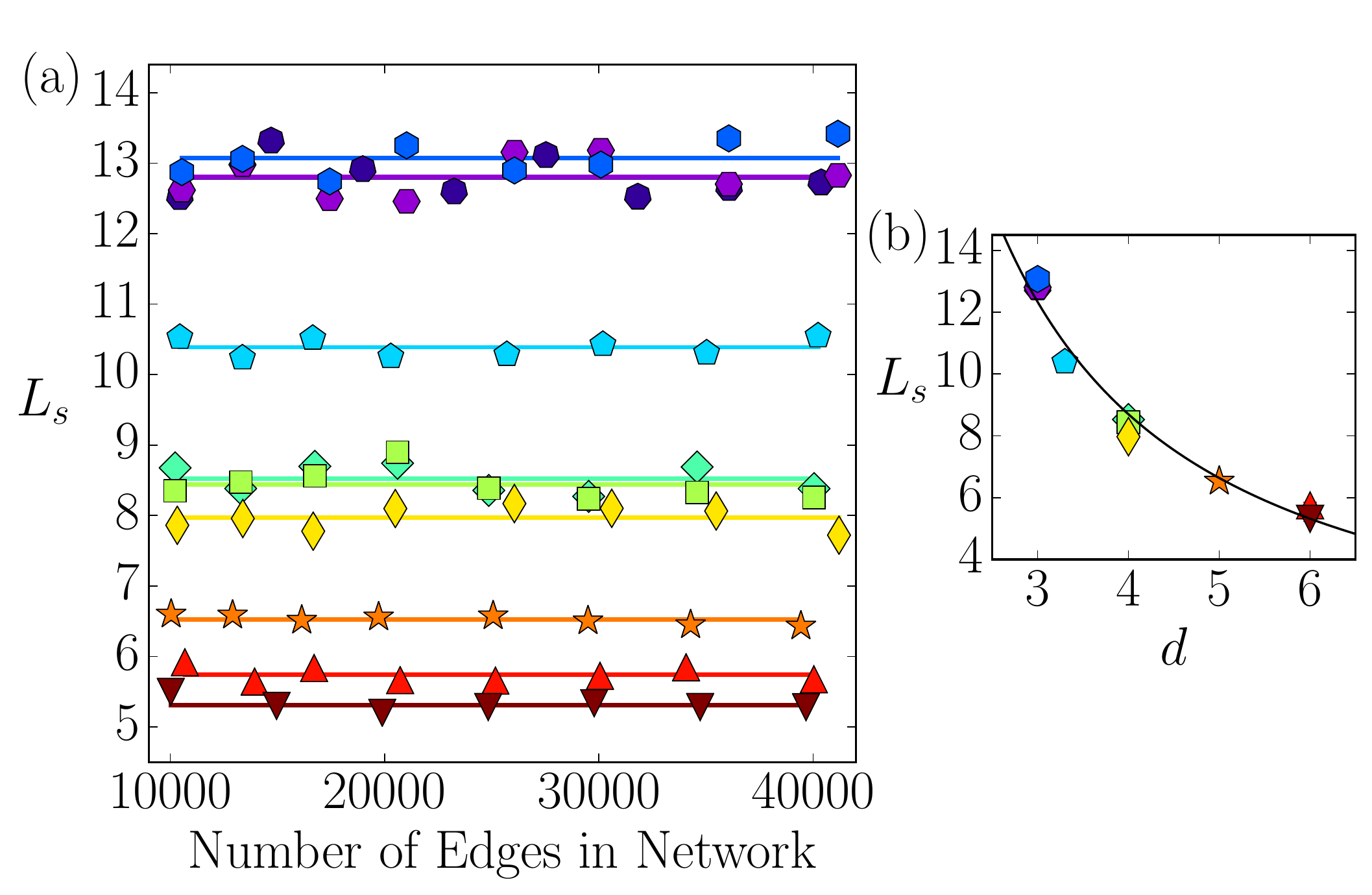} 
\caption{(a) Shielding length $L_s$ as a function of network size for a variety of networks shown in Fig.~\ref{lattices}. For a fixed network type, $L_s$ remains constant as the network grows in number of nodes. Solid lines indicate the mean $L_s$ for a network type. (b) $L_s$ is a monotonically decreasing function of the mean network degree.}
\label{SL_topology}
\end{center}
\end{figure}

In Fig.~\ref{SL_topology}(a) we show that for a single network type, $L_s$ is independent of network size above a certain threshold size. For each network of the same type and a sufficiently large size, $L_s$ is closely approximated by the mean $L_s$ over all network sizes, indicated by the solid lines. First of all, this confirms that finite boundary effects do not influence $L_s$ for these network sizes. This is expected because boundary effects are seen on the order of $L_s$, so finite size effects are negligible for networks larger than a few shielding lengths.

Moreover, we find that $L_s$ is a strictly decreasing function of the mean node degree $d$. We plot the mean $L_s$ for each network type as a function of the mean network degree in Fig.~\ref{SL_topology}~(b). This relation can be fit by $L_s\sim d^{-1.21}$ for the limited range of available data. Thus, network degree dictates $L_s$, as expected since $d$ is a measure of the network connectedness, strongly correlated with other measures such as the effective resistance between neighboring nodes. This result indicates that networks which are more tightly connected require a smaller area to reroute displaced flow, and inversely, the shielding effects of a vein drop off faster in a network with higher $d$. Because the shielding length is dictated by the average node degree (and thus the network connectivity) and not the main vein conductance, it can be classified as a primarily topological effect.

\subsection{Interactions of Multiple Veins}
\label{sec:multiple_veins}
\begin{figure}[h]
\begin{centering}
\hspace*{-.5cm}
\begin{tabular}{ccc}
\includegraphics[height=29mm]{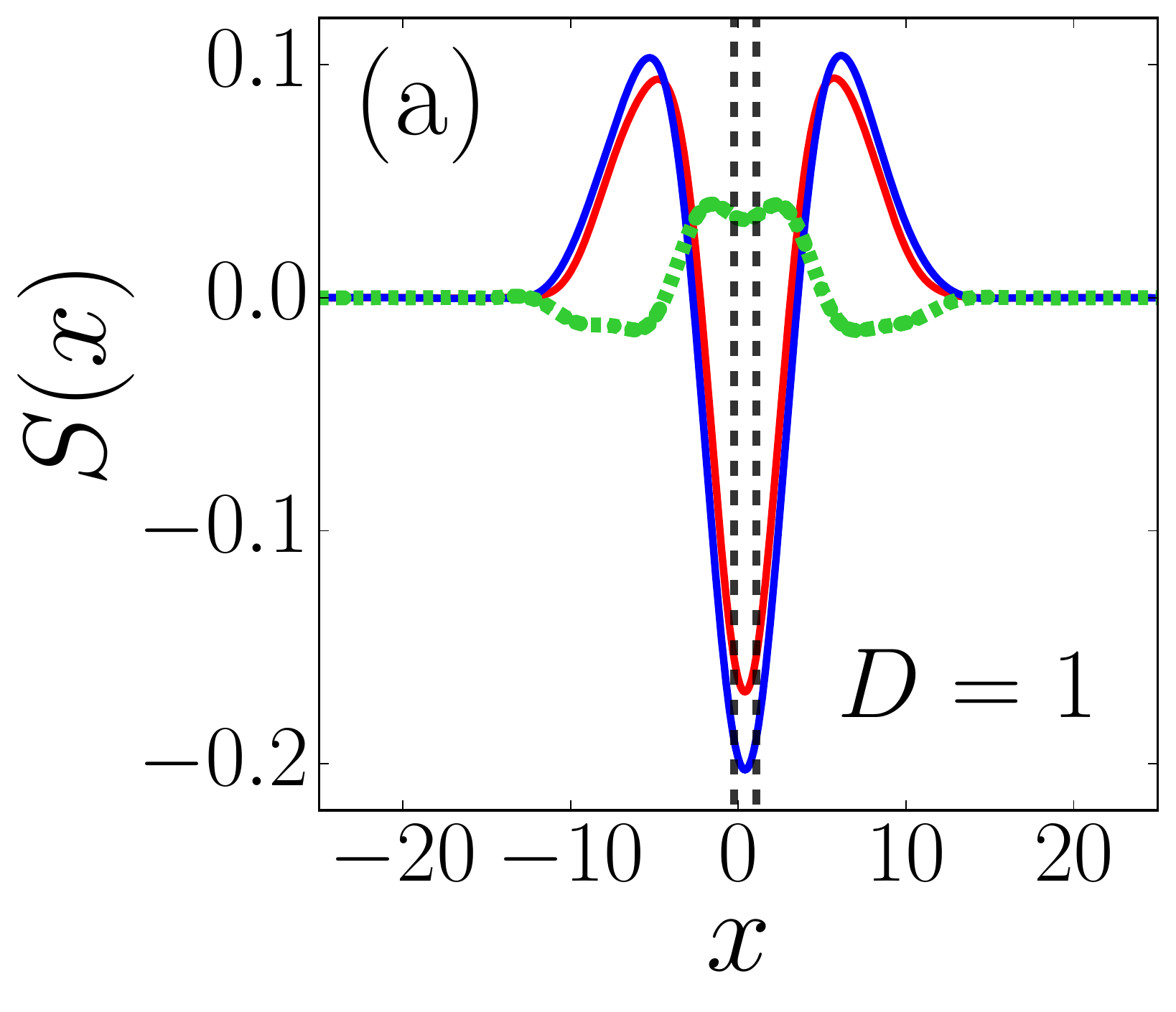} &
\includegraphics[height=29mm]{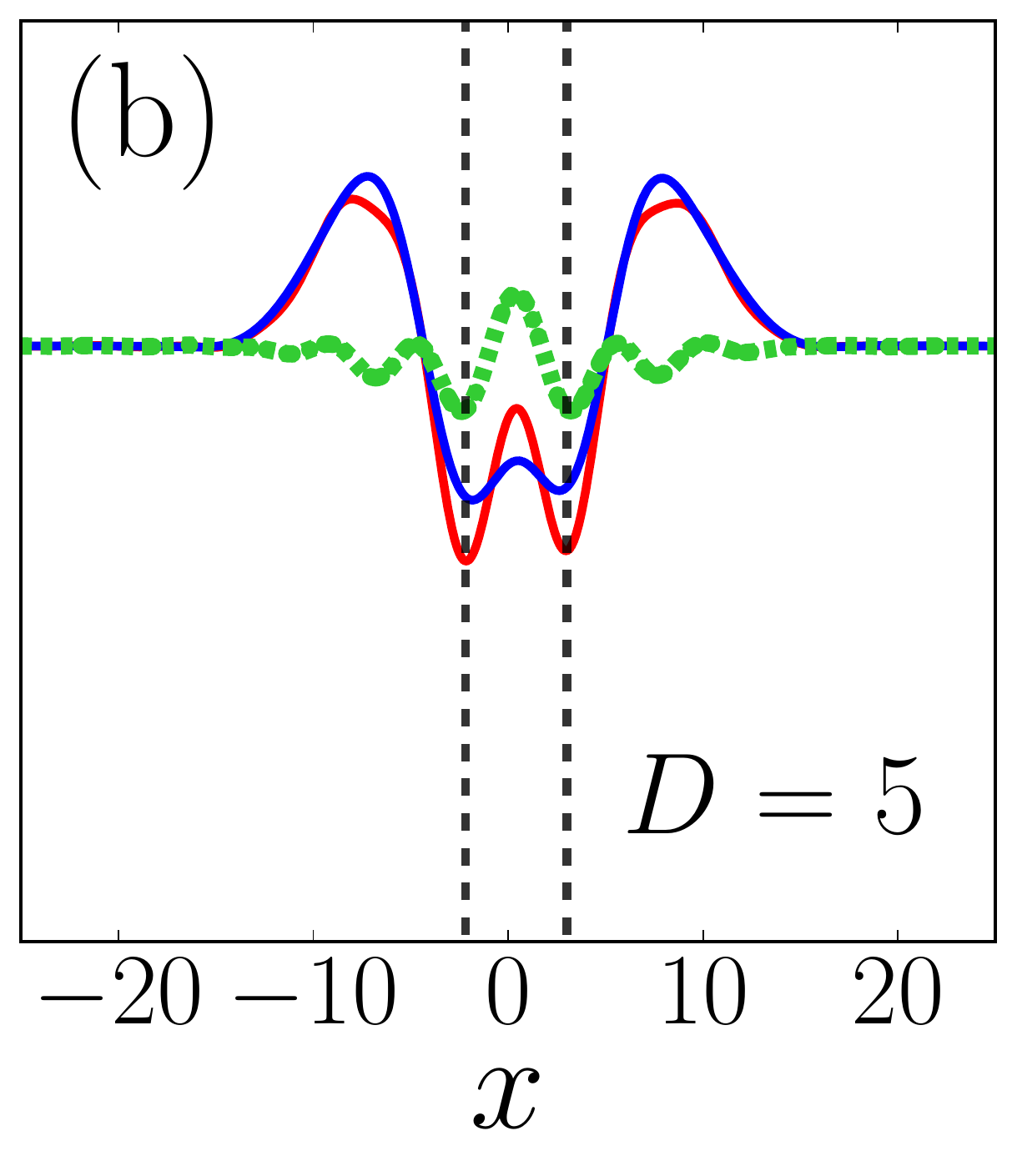} &
\includegraphics[height=29mm]{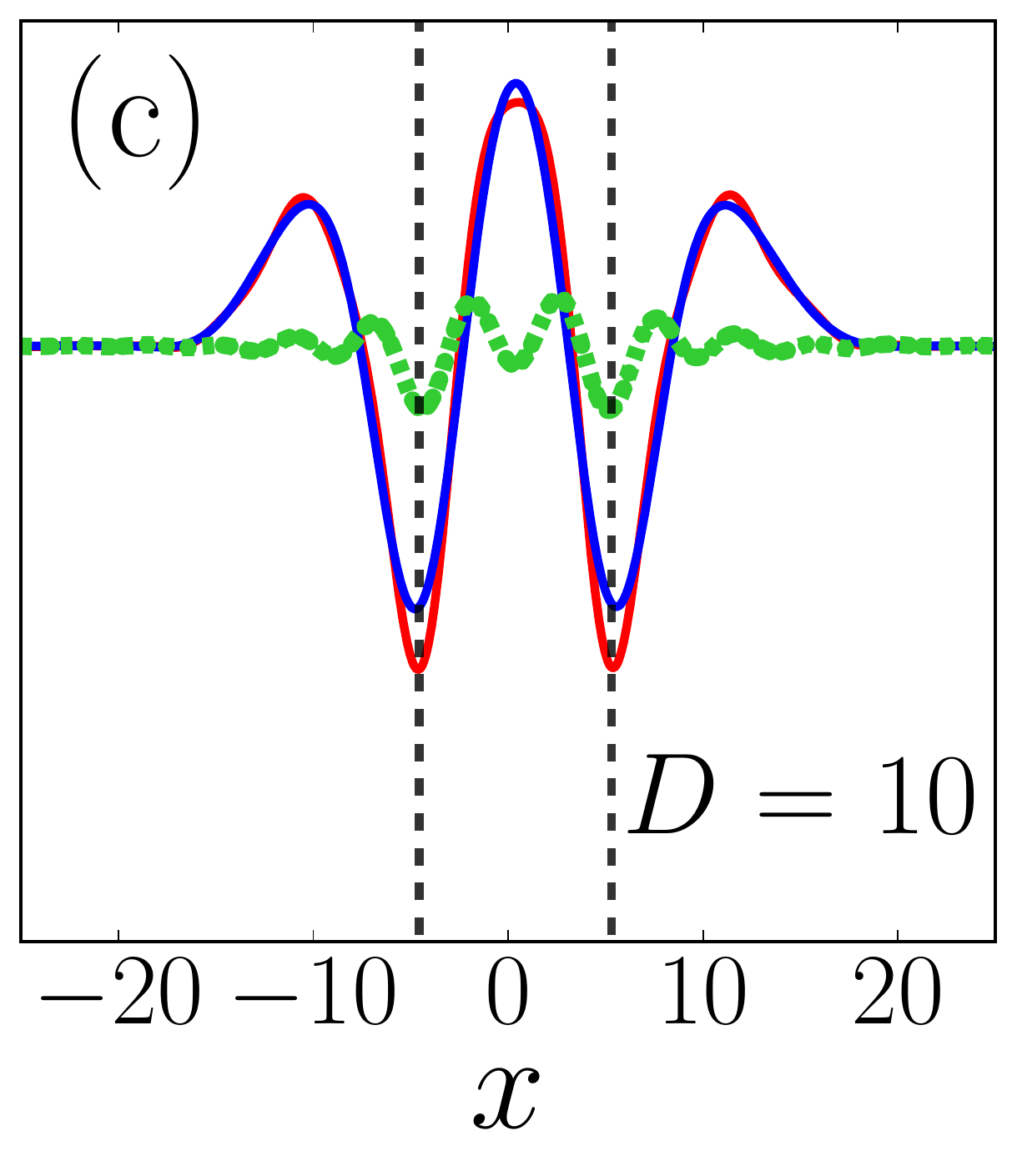} \\
\end{tabular}
\caption{Edge shielding fit function $S(x)$ for three RP triangular tilings of 6000 nodes with bulk edge conductance 1 and two veins of conductance 5 at varying separation $D$, indicated by the vertical dashed lines. The composite system $S(x)$ (solid red curve) is compared to the sum of two single-vein systems $S_{\text{L}}(x) + S_{\text{R}}(x)$ (solid blue curve). The residual is $\Delta S = S(x) - S_{\text{L}}(x) - S_{\text{R}}(x)$ (dotted green curve).}
\label{two_veins}
\end{centering}
\end{figure}

To describe the effects of complex vein hierarchies on network resilience, we begin by quantifying the interaction of edge shielding fit functions for two nearby veins. To see if the edge shielding is an additive effect, we compare $S(x)$ for a system of two veins with separation $D$ with the sum $S_{\text{L}}(x) + S_{\text{R}}(x)$ from two distinct systems, one with just the left vein present and one with the right vein present. The amplitude of the residual signal, $\Delta S = S(x) - S_{\text{L}}(x) - S_{\text{R}}(x)$, is a measure of nonlinearity in the system: if the system with two veins is exactly a sum of the shielding effects from two separate veins, the residual will be zero. We plot $\Delta S$ for two veins of increasing separation $D$ in Fig.~\ref{two_veins} and we find that it becomes zero for $D \ge 12$. The shielding length for an individual vein has been found to be 5.6, so two veins become independent when their shielding lengths no longer overlap. Even for small separation distance $D$, the residual $\Delta S$ is small relative to $S(x)$.

Taking inspiration from natural hierarchically ordered networks, we examine systems of multiple veins with two different hierarchies. Leaf venation networks often feature complex hierarchical structures, the purpose of which is not completely known \cite{Sack2012}. Brain vasculature contains penetrating arterioles and venules that form a class of hierarchy, but interestingly these higher order veins do not have loops; loops exist in the denser smaller veins that form the bulk region. 

Here we want to explore how presence of intersecting higher order veins affects resilience. We compare a hierarchy of strictly vertical veins (parallel hierarchy) with a hierarchy that has both vertical and horizontal veins arranged in a grid (grid hierarchy). As a null model comparison we use a network with edges chosen at random to be highly conductive (null hierarchy), which lacks any kind of hierarchical ordering. We generate networks with these three types of hierarchies at different values of vein density to see if there is an favorable design for resilience. 


We define the occupation fraction $f$ of a network with veins to be the fraction of on-vein edges to bulk edges. Although any $0<f<1$ is allowed in principle, it is limited by the fraction of vertical edges in the underlying lattice. For a square grid, half of the edges are oriented vertically, so the maximal occupation fraction is $f=0.5$. We find that for the RP triangular tiling $f \sim 0.33$ is the highest possible occupation fraction with non-intersecting veins. For the grid hierarchy, where intersection of perpendicular veins is allowed, $f \sim 0.6$ is a reasonable upper limit for the RP tiling. To generate networks of higher $f$, we invert the edge conductances for a lower occupation fraction network. For example, a network with $f = 0.8$ is generated by taking the network with $f = 0.2$ and switching every edge with conductance 1 to conductance 5 and vice versa. The resulting network has thicker veins consisting of several columns of edges, interspersed with small strips with conductance 1.

\begin{figure*}
\includegraphics[width=180mm]{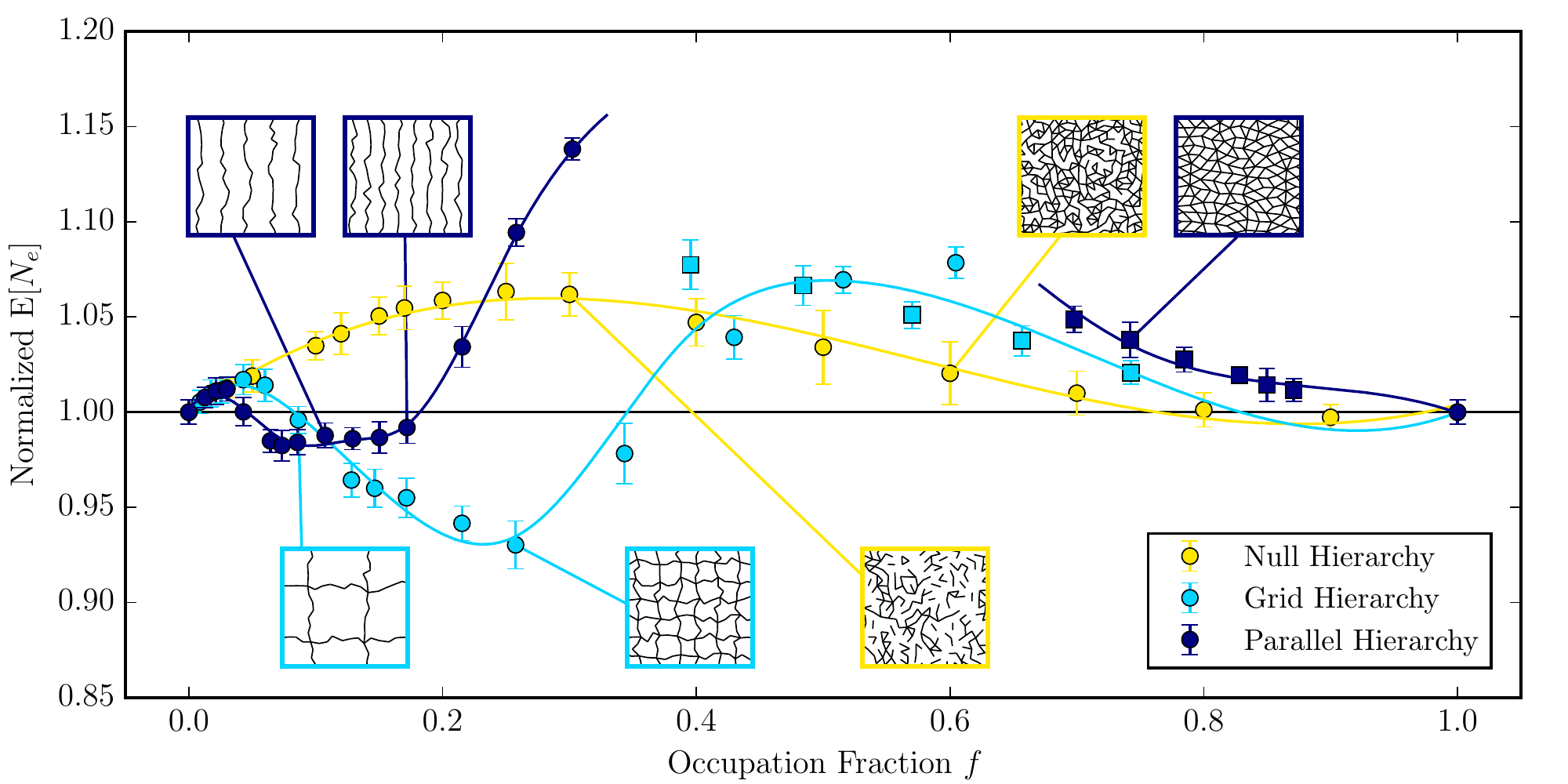}
\caption{The expected number of edges in the damage zone as a function of vein occupation fraction $f$ for three different types of vein hierarchies. Networks are RP triangular tilings with 6000 nodes with bulk edge conductance 1 and vein edge conductance 5. Each data point is the average over 10 different tilings; error bars denote 5 standard deviations across the 10 RP tilings. Circles indicate networks formed by adding veins and squares indicate networks formed by inverting the conductances of a network with occupation fraction $1-f$. Values of $\text{E}[N_e]$ are normalized with respect to a network with no veins, so $\text{E}[N_e]>1$ indicates that a network is less resilient than the veinless network and $\text{E}[N_e]<1$ indicates a more resilient network. For the null hierarchy, increasing $f$ always yields a less resilient network. For the parallel and grid hierarchies, adding more veins first increases the resilience of the network by providing a shielding effect to off-vein edges, but then decreases resilience by filling the network with edges that have a high damage cost. A spline fit is provided to guide the eye for type of vein hierarchy. The central piece of the curve fit for the parallel veins has been removed to avoid extrapolating the location of the maximum due to a sparsity of data in that region. Insets are partial network segments $4\%$ of the total network area in size, shown to illustrate the vein hierarchies of different occupation fraction.}
\label{efficiency}
\end{figure*}

We consider the network resilience for a class of vein hierarchies as a function of $f$. To quantify network resilience, we use $\text{E}[N_e]$, the expected number of edges in the damage zone for an edge, using the distribution of $N_e$ across all edges in the network. This is an estimate for the expected amount of disruption in the network if one edge is damaged with uniform probability over all edges. A high value of $\text{E}[N_e]$ means that on average, a large amount of edges will exceed their threshold value of flow disruption, and thus a larger proportion of the network will have disrupted functionality in the event of an occlusion. A network with high $\text{E}[N_e]$ is more sensitive to damage and thus less resilient.

Figure~\ref{efficiency} shows how $\text{E}[N_e]$ changes as a function of $f$ for the three hierarchy types. For a single network with a given hierarchy and $f$, we fit the distribution of $N_e$ to a Gaussian function and record the expected value. We repeat the calculation for 10 different instances of the RP tiling for each value of $f$. Data points represent the average $\text{E}[N_e]$ and error bars indicate five standard deviations of $\text{E}[N_e]$ across different instantiations of the 10 RP tilings. One standard deviation due to changes in the underlying lattice is less than 0.5 percent for each data point, which is negligible compared to the magnitude of variation of $\text{E}[N_e]$. The y-axis is scaled by $\text{E}[N_e]$ for a veinless network, so $\text{E}[N_e]$ of the initial $f=0$ network is 1. Note that as $\text{E}[N_e]$ decreases the network becomes more resilient, since the deletion of one edge results in a smaller number of disrupted edges than in the uniform veinless network. 

The first thing to note is that $\text{E}[N_e] = 1$ for $f = 1.0$, so the network where every edge is a highly conductive vein has  the same resilience as the veinless network. This is expected as the threshold expression was designed to capture the effects of hierarchy and not the absolute value of the conductance. Changing the conductance of every edge in the network results in scaling all $C_{ij}$ by 5 and all $L_{ij}^{-1}$ by {1/5} in Eq.~\ref{env_condition}. This factor cancels, resulting in the same threshold expression, so the damage zone of an edge will stay the same if all network conductances are rescaled by the same constant.

We have shown that the presence of a vein increases the resilience of some edges but decreases the resilience of others, however it is not obvious which of these effects is dominant. As seen in the form of $S(x)$, plotted in Fig.~\ref{shield}, edges within the shielding length typically have a lower $N_e$ and edges just outside the shielding length, as well as edges that are on the actual vein, have a higher $N_e$. As highly conductive veins are added to the system, edges close to the vein will experience a shielding effect, increasing the resilience of the system, and edges in a strip further away will contribute to the decreasing resilience of the system. We suspect that once the venation attains the density such that every edge is within one shielding length of the vein the system will reach maximum resilience.

We find that networks of the three hierarchies exhibit different behavior in their global resilience as $f$ varies. The null hierarchy network attains a single shallow maximum at $f = 0.28$, or 28 percent occupancy. Because $\text{E}[N_e] > 1$ over the entire range of $f$, the null hierarchy is always less resilient than a network with no veins. This can be explained by reasoning that the shielding effect holds only when there is a nearly continuous vein present. The two networks with hierarchical vein structure first attain a minimum value before reaching a maximum. The maximum of the grid hierarchy occurs at $f = 0.50$. Because the parallel hierarchy is not well-defined around its maximum value, we do not extrapolate the exact value. The minima of the parallel hierarchy and the grid hierarchy occur at $f = 0.10$ and $f = 0.23$, respectively. For the parallel hierarchy, this value of $f$ is reasonable for natural flow networks. An example be found in leaf venation, as demonstrated in Appendix~\ref{appendix:real_networks}.

We interpret the minimum as the occupation fraction that produces the most resilient network. For $f < 0.05$, $\text{E}[N_e]$ for the two hierarchies behaves almost identically; in this regime the veins are too sparse to have a meaningful impact on the global resilience. For $f > 0.1$, the grid hierarchy is always more resilient than the parallel hierarchy. The parallel hierarchy is only able to provide a 2 percent decrease of $\text{E}[N_e]$ compared to a network with no veins, while the grid hierarchy is able to provide a 7 percent decrease. The position of the minimum is determined by the competition between the positive and negative effects of adding a highly conductive vein. Increasing the vein density increases resilience to some degree, but since damage of on-vein edges results in a high amount of displaced current, soon veins become detrimental to the overall network resilience. 

The position of the maximum indicates the vein density generating the least resilient network. A network that is minimally resilient is maximally sensitive in the sense that the damage response is not localized, and that a distant edge is likely to detect that damage has occurred. This may be a useful feature for some applications; for instance, if the network has the ability to mediate damage by adding edges it may be beneficial to measure that damage has occurred far from the damage site. In this case, the parallel vein hierarchy is preferable. 

Note that most likely actual biological flow networks are optimized for many functions, and resilience potentially plays a minor part in network design. The two types of hierarchy explored are just to offer an intuitive idea about the effects of hierarchical architecture in network resilience, and not meant as an exhaustive optimization.

\section{Summary and Conclusions}\label{discussion}

The ability of networks to withstand damage with limited consequences to their function is important for understanding biological networks and for designing engineered networks. The question of how a network reroutes the flow in the event of an occlusion and how the hierarchical vascular architecture determines the size of the affected areas has been studied empirically in the context of ischemic strokes \cite{Nishimura2010} but no significant theoretical exploration has to our knowledge taken place. A theoretical framework of the effects of topology and hierarchy in flow displacement after an occlusion would allow a more fundamental understanding of why some vascular architectures are more susceptible to damage than others. For this reason, in this work we have studied the resilience of flow networks by examining how displaced flow is distributed throughout the network after perturbative damage. We have shown that network hierarchy has resounding implications for network resilience. In particular, we found that the presence of a vein in a network changes the resilience by providing an efficient channel to reroute displaced flow.

We have developed a local and global measure of network resilience. The damage zone caused by edge removal tracks the network area that has experienced a significant disruption after an edge occlusion. By separately analyzing different parts of the damage zone, we can understand the underlying mechanism of shielding that the vein provides to its surrounding edges. Further, this can be turned into a global network measure by considering how the ensemble of damage zones changes across networks of different architectures. We believe that the damage zone is a biologically meaningful measure, as it can represent an area of tissue that has suffered hypoxia after a stroke.

We find that a highly conductive vein contains the spread of flow disruption for damage to edges close to the vein, but increases the effect of damage on edges further away. We call the change in damage zone the edge shielding, since the vein tends to decrease the damage zone for the nearest edges. Specifically, we show that the vein serves to prevent displaced flow due to damage on one side of the vein from reaching the other side. We have characterized the length scale of the vein effect through the shielding length $L_s$, i.e. the distance from the vein at which vein effects start to decay.

We have shown that $L_s$ for a network is primarily controlled by the network topology. In particular, increasing the conductance of the vein beyond the initial thickness necessary to establish a shielding length does not significantly change $L_s$. If $C_{\text{vein}}$ is the ratio of the vein conductance to the ambient conductance, we find that $C_{\text{vein}} = 1.5$ is sufficient to observe a shielding effect and any $C_{\text{vein}} > 5$ results in $L_s = 5.6$. By comparing uniform networks with a variety of topologies, we have shown that the shielding length is determined by the mean network degree, with more tightly connected networks having a smaller $L_s$. Because the shielding length is primarily affected by the network wiring and has a slight but much less weaker dependence on the conductance of the central vein, we classify the shielding length as a primarily topological effect.

Lastly, we have used the intuition acquired by studying a single vein to analyze networks with varying vein hierarchies. The shielding effects are nonlinear, and when two or more veins are present, their effects are coupled. We have shown that veins separated by two shielding lengths affect the displaced current independently. We study two types of vein hierarchies, one with the veins arranged in parallel tracks, and one where they form a grid, and compare with a null hierarchy that has no continuous veins, where edges are chosen at random to have high conductance. We find that the null hierarchy can only increase the network resilience. However, the grid venation network is able to increase the global network resilience by 7 percent, while the parallel network is only able to provide a 2 percent effect.  



The shielding length is an inherently discrete effect. Consider the limit in which the number of network nodes grows while the network is contained in a square box of length $L$, so the mean edge length $a = L/\sqrt{N}$. For a fixed network topology, $L_s$ is proportional to $a$ regardless of $N$, as seen in Fig.~\ref{SL_topology}. As $N$ grows to infinity, $L_s$ becomes zero, and the shielding effect of the vein disappears. Thus the shielding property is a truly discrete effect, lacking a continuous limit. 


If indeed resilience is a feature that biological networks favor, then this should be reflected in certain network features. We have shown that from a damage perspective, a greater investment in network resources to build and maintain a vein will not necessarily yield more benefits. This is seen in two ways: increasing the vein conductance will not always yield a greater shielding length, and increasing the vein density will not always result in greater network resilience. This leads to the notion that there is an optimal vein structure that balances the cost and benefits to the entire network. In certain circumstances it might be beneficial to have large damage zones (which would translate to low resilience), as this would spread out the displaced current.

For ease in visualizations and computation, in this work we chose to focus on planar networks. From preliminary work on 3D networks, we expect that some of our results hold in non-planar networks, whereas others, like the size of the negative $S(x)$ zone near the vein, are dimensionality dependent. The full study of the 3D system is complex and beyond the scope of this work, and thus reserved for a future publication. 

While this model was meant as an initial step towards visualizing and understanding the role of highly conductive veins in flow rerouting for biological systems, real biological networks have features that are not captured by the model, yet may play an important role in network optimization and resilience. One implicit assumption is that a single edge is able to sustain any amount of flow, even independent from the initial edge flow. However, a biological system will have limits on the node pressures that it cannot exceed without breaking the connections of the network. We do not consider these limits, but it would be interesting to see how the extent of damage would change if there were an imposed limit on the pressure drop across an edge. In addition, it is known that brain vasculature is adaptive: network edges are able to dilate or contract their ambient diameter in order to modulate their conductance in response to changes in flow \cite{Shih2015}. This ability has strong consequences for the flow redistribution, and our work could shed light on the extent of vascular remodeling after network injury.

Despite the focus on vascular networks, this work is more general and should apply to any laminar flow or electrical network. We have made the initial steps for a mechanistic statistical analysis of the effects of damage in flow networks. Future work should be devoted in understanding the specifics of more complex network architectures and  response.

\begin{acknowledgments}
TG would like to thank L. Papadopoulos for insightful discussion during this work. We also thank M. Ruiz-Garc\'ia and B. G. Chen for a thorough reading of the manuscript. We are grateful to Tarmily Wen, who processed the leaf sample. This work was partially supported by the NSF Award PHY-1554887, the Burroughs Wellcome Fund and by NSF through the University of Pennsylvania Materials Research Science and Engineering Center (MRSEC) (DMR-1720530). EK would like to acknowledge support from the American Institute of Mathematics through the SQuaRE research program.
\end{acknowledgments}



\appendix\section{Removing a single edge is equivalent to adding a single flow dipole} \label{appendix:dip_calc}

For a general flow network, the known quantities are the edge conductances $C_{ij}$ and the injected or extracted node currents $Q_i$. We will solve for the node potentials $v_i$ and the edge currents $I_{ij}$. For the system to be physical, the total node current must be conserved: $\sum_i Q_i= 0$. From Ohm's Law and Kirchhoff's vertex law
\begin{equation}
\begin{aligned}
&I_{ij} = C_{ij}(v_i - v_j) \\
&\sum_j I_{ij} = Q_{i}
\end{aligned}
\label{ohm}
\end{equation}
we derive the basic flow equation
\begin{equation}
\begin{aligned}
Q_i &=  \sum\limits_j C_{ij} (v_i - v_j)
= \sum_j \delta_{ij}v_j \sum\limits_n C_{in} - \sum\limits_j C_{ij}v_j \\
&=  \sum\limits_j (\delta_{ij}\sum\limits_n C_{in} - C_{ij})v_j = \sum_j L_{ij} v_j
\end{aligned}
\end{equation}
since $\delta_{ij}\sum\limits_nC_{in} - C_{ij}$ is the $ij^{th}$ entry of the weighted graph Laplacian $\overline{\overline{L}}$. Thus we have 
\begin{equation}
\begin{aligned}
\overline{Q} = \overline{\overline{L}}\overline{v}
\end{aligned}
\end{equation}
and we can take the pseudo-inverse of the Laplacian to get $\overline{v} = \overline{\overline{L}}^{-1}\overline{Q}$. Using \ref{ohm} we can calculate the edge flow:
\begin{equation}
\begin{aligned}
I_{ij} = C_{ij} \bigg( (L^{-1}Q)_i - (L^{-1}Q)_j \bigg)
\end{aligned}
\end{equation}
Suppose that we change the graph by perturbing the conductance of edge $\kappa\lambda$, so that $C'_{\kappa\lambda} = C_{\kappa\lambda} + \delta C$. The new graph conductances thus read:
\begin{equation}
C'_{ij} = C_{ij} + \delta C\delta_{ij, \kappa\lambda}
\end{equation}
The perturbation to the Laplacian is a rank-1 matrix, namely
\begin{equation}
\begin{aligned}
\overline{\overline{L}}' = \overline{\overline{L}} + uu^T, \quad \text{where} \, u_i = \sqrt{\delta C}(\delta_{i\kappa} - \delta_{i\lambda}) \\
\end{aligned}
\end{equation}
To find the inverse of the perturbed Laplacian, we use the Sherman-Morrison formula:
\begin{equation}
\begin{aligned}
& (L_{ij}')^{-1} = (L_{ij} + uu^T)^{-1} \\
&= L_{ij}^{-1} - \bigg(  \frac{L^{-1}uu^TL^{-1}}{1 + u^TL^{-1}u}  \bigg)_{ij} \\
\end{aligned}\label{ShermanMor}
\end{equation}
We can rewrite $u^TL^{-1}u$ in terms of the effective resistance $R^{\text{eff}}_{\kappa\lambda}$ between nodes $\kappa$ and $\lambda$. The effective resistance (or resistance distance) between two nodes in a graph is defined as the resistance of the system when a test current $I_{\text{test}}$ is injected in $\kappa$ and extracted from $\lambda$:
\begin{equation}
\begin{aligned}
R^{\text{eff}}_{\kappa\lambda} &= \frac{v_\kappa - v_\lambda}{I_{\text{test}}} \\
&=  \frac{1}{I_{\text{test}}} \sum_n ( L_{\kappa n}^{-1} - L_{\lambda n}^{-1} ) Q_n \\ 
&= \frac{1}{I_{\text{test}}} \sum_n ( L_{\kappa n}^{-1} - L_{\lambda n}^{-1} ) (
	\delta_{\kappa n}I_{\text{test}} - \delta_{\lambda n} I_{\text{test}}) \\ 
& = L^{-1}_{\kappa\kappa} -  L^{-1}_{\kappa\lambda} -  L^{-1}_{\lambda\kappa} +  L^{-1}_{\lambda\lambda} \\
&= \frac{1}{\delta C}u^TL^{-1}u
\end{aligned}
\end{equation}
So $1 + u^TL^{-1}u = 1 + \delta C R^{\text{eff}}_{\kappa\lambda}$. Let $\Omega \equiv 1 + \delta C R^{\text{eff}}_{\kappa\lambda}$, which is an $ij$-independent constant. Then from Eq.~\ref{ShermanMor}
\begin{equation}
\begin{aligned}
L'^{-1}_{ij} &= L_{ij}^{-1} - \frac{1}{\Omega}(L^{-1}uu^TL^{-1})_{ij} \\
&= L_{ij}^{-1} - \frac{1}{\Omega}(L^{-1}u)_i(L^{-1}u)^T_j
\end{aligned}
\end{equation}
Evaluating:
\begin{equation}
\begin{aligned}
(L^{-1}u)_i &= \sum_{j}L_{ij}^{-1} u_j =  \sqrt{\delta C}(L_{i\kappa}^{-1} - L_{i\lambda}^{-1})
\end{aligned}
\end{equation}
\begin{equation}
\begin{aligned}
(L_{ij}')^{-1} &=
L_{ij}^{-1} - (L^{-1}u(L^{-1}u)^T)_{ij} \\
&= L_{ij}^{-1} - \frac{\delta C}{\Omega} (L_{i\kappa}^{-1} - L_{i\lambda}^{-1})(L_{\kappa j}^{-1} - L_{\lambda j}^{-1})
\end{aligned}
\end{equation}
The edge current after the perturbation is given by
\begin{equation}
\begin{aligned}
I'_{ij} = C'_{ij}(v'_i - v'_j)
\end{aligned}
\end{equation}
Evaluating:
\begin{equation}
\begin{aligned}
& v'_i - v'_j = \sum\limits_{m}(L')_{im}^{-1} Q_m - \sum\limits_{m}(L')_{jm}^{-1} Q_m \\
& =  \sum \limits_m \bigg( 
	L_{im}^{-1} - \frac{\delta C}{\Omega} (L_{i\kappa}^{-1} - L_{i\lambda}^{-1})
	(L_{\kappa m}^{-1} - L_{\lambda m}^{-1}) \\
&\qquad-L_{jm}^{-1} + \frac{\delta C}{\Omega} (L_{j\kappa}^{-1} - L_{j\lambda}^{-1})
	(L_{\kappa m}^{-1} - L_{\lambda m}^{-1})   \bigg) Q_m\\
& =	\sum\limits_m(L_{mi}^{-1}  - L_{mj}^{-1})Q_m \\
&\qquad - \frac{\delta C}{\Omega} \Lambda_{ij\kappa\lambda} 
		\sum\limits_m (L_{m\kappa}^{-1} - L_{m\lambda}^{-1}) Q_m \\
& = \frac{I_{ij}}{C_{ij}} - \frac{\delta C}{\Omega} \Lambda_{ij\kappa\lambda} \frac{I_{\kappa\lambda}}{C_{\kappa\lambda}}
\end{aligned}
\end{equation}
where $\Lambda_{ij\kappa\lambda} = L_{\kappa i}^{-1} - L_{\lambda i}^{-1} - L_{\kappa j}^{-1} + L_{\lambda j}^{-1}$.
The change in current before and after bond $\kappa\lambda$ is broken reads:
\begin{equation}
\begin{aligned}
&\Delta I_{ij} =I'_{ij} - I_{ij} = \\
& = (C_{ij} + \delta C\delta_{ij, \kappa\lambda})
	\bigg( \frac{I_{ij}}{C_{ij}} - 
	\frac{\delta C}{\Omega} \Lambda_{ij\kappa\lambda} \frac{I_{\kappa\lambda}}{C_{\kappa\lambda}} \bigg) - I_{ij} \\
& = \delta_{ij, \kappa\lambda}
	\bigg( \frac{\delta C I_{ij}}{C_{ij}} - 
	\frac{\delta C^2}{\Omega} \Lambda_{ij\kappa\lambda} \frac{I_{\kappa\lambda}}{C_{\kappa\lambda}} \bigg) - \frac{\delta C}{\Omega}\frac{C_{ij}}{C_{\kappa\lambda}}I_{\kappa\lambda}\Lambda{ij\kappa\lambda}\\
\end{aligned}\label{deltaIJ}
\end{equation}
We can now rephrase the problem slightly. Suppose that we have the original network with a different set of current sources and sinks: let $Q^{(\kappa\lambda)}$ be the set of node currents such that a current of magnitude $I_{\kappa\lambda}$ (the current flow through edge $\kappa\lambda$ in with the original $Q_i$) is injected at node $\kappa$ and extracted at node $\lambda$. Thus, $Q_i^{(\kappa\lambda)} = I_{\kappa\lambda}(\delta_{i\kappa} - \delta_{i\lambda})$. We can then write down the basic flow equations for the system with the new dipole current source and sink but the original graph Laplacian: 
\begin{equation}
\begin{aligned}
&v_i^{(\kappa\lambda)} - v_j^{(\kappa\lambda)} = 
	(L^{-1}Q^{(\kappa\lambda)})_i - (L^{-1}Q^{(\kappa\lambda)})_j \\
	& = \sum_m L^{-1}_{im}Q^{(\kappa\lambda)}_m -  \sum_m L^{-1}_{jm}Q^{(\kappa\lambda)}_m\\ 
	& = I_{\kappa\lambda} \bigg( \sum_m L^{-1}_{im} (\delta_{m\kappa} - \delta_{m\lambda}) 
	- \sum_m L^{-1}_{jm} (\delta_{m\kappa} - \delta_{m\lambda}) \bigg) \\
	&= I_{\kappa\lambda} (L_{i \kappa}^{-1} - L_{i \lambda}^{-1} - L_{j \kappa}^{-1} + L_{j \lambda}^{-1}) 
	= I_{\kappa\lambda} \Lambda_{ij\kappa\lambda}
\end{aligned}\label{vkappalam}
\end{equation}
and, as an analogue to equation \ref{ohm}:
\begin{equation}
\begin{aligned}
	I_{ij}^{(\kappa\lambda)} &= C_{ij} (v_i^{(\kappa\lambda)} - v_j^{(\kappa\lambda)})
\end{aligned}
\label{ohm2}
\end{equation}
where the superscript $(\kappa\lambda)$ denotes quantities evaluated with the node currents $Q^{(\kappa\lambda)}$. 
Combining Eq.~\ref{vkappalam} and Eq.~\ref{ohm2} gives:
\begin{equation}
\begin{aligned}
\Lambda_{ij\kappa\lambda} = \frac{v_i^{(\kappa\lambda)} - v_j^{(\kappa\lambda)}}{I_{\kappa\lambda}}
	=\frac{I_{ij}^{(\kappa\lambda)}}{ C_{ij} I_{\kappa\lambda}}
\end{aligned}
\end{equation}
In the case where the edge is completely removed, $\delta C = - C_{\kappa\lambda}$. So for $ij \ne \kappa\lambda$, Eq.~\ref{deltaIJ} reads:
\begin{equation}
\begin{aligned}
	\Delta I_{ij} = \frac{1}{\Omega} I_{ij}^{(\kappa\lambda)} = 
    \frac{1}{1 - C_{\kappa\lambda}R_{\kappa\lambda}^{\text{eff}}}I_{ij}^{(\kappa\lambda)}
\end{aligned}
\end{equation}
This shows that for all edges besides $\kappa\lambda$, the displaced edge current $\Delta I_{ij}$ in a network after removing edge $\kappa\lambda$ is proportional to the edge current through $I_{ij}^{(\kappa\lambda)}$ with the undamaged structure but new node currents $Q^{(\kappa\lambda)}$. 

\section{The damage zone is independent of the initial currents and the current sources and sinks} \label{appendix:dz_indep}

Using Eq.~\ref{vkappalam} and Eq.~\ref{ohm2}, we write $I_{ij}^{(\kappa\lambda)} = C_{ij} I_{\kappa\lambda} (L_{i \kappa}^{-1} - L_{i \lambda}^{-1} - L_{j \kappa}^{-1} + L_{j \lambda}^{-1})$. Then, 
\begin{equation}
\bigg| \frac{\Delta I_{ij}}{I_{\kappa\lambda}}\bigg| = 
\bigg| \frac{I_{ij}^{(\kappa\lambda)}}{\Omega I_{\kappa\lambda}} \bigg|=
\bigg| \frac{C_{ij}(L_{i \kappa}^{-1} - L_{i \lambda}^{-1} - L_{j \kappa}^{-1} + L_{j \lambda}^{-1})}{1-C_{\kappa\lambda}(L^{-1}_{\kappa\kappa} - 2L^{-1}_{\kappa\lambda} + L^{-1}_{\lambda\lambda})} \bigg|
\end{equation}
Edge $ij$ is included in the damage zone for edge $\kappa\lambda$ with threshold $t$ if $|\frac{\Delta I_{ij}}{I_{\kappa\lambda}}| > t$, so the equivalent condition is:
\begin{equation}
 	\bigg| \frac{ C_{ij} (L^{-1}_{i\kappa} - L^{-1}_{i\lambda} - L^{-1}_{j\kappa} + L^{-1}_{j\lambda})}
	{1 - C_{\kappa\lambda}(L^{-1}_{\kappa\kappa} - 2L^{-1}_{\kappa\lambda} + L^{-1}_{\lambda\lambda})} \bigg| > t
\end{equation}
which is independent of the initial edge currents. 
\section{Network and Vein Generation} \label{appendix:network}
We use disordered networks in simulations to emulate biological networks and to avoid lattice effects observed in periodic networks. The most common network used is the randomly packed (RP) triangular tiling. To create this tiling, we first generate a set of Poisson distributed points on a square domain and apply a repulsive pointwise potential iteratively to generate a set of randomly but uniformly distributed points. These points are used for the network nodes. Edges are formed from the Delaunay triangulation of the nodes. The final network has an approximately uniform distribution of edge lengths. Most commonly, the conductances of all edges are set to be a constant equal to 1 in our dimensionless units. For some applications, we set edge conductances to be inversely scaled with length and normalized by the mean edge length, which  yields a distribution of edge conductances centered at 1. 

Veins are formed by selecting a subset of edges from the underlying network to set to a high conductance, equal to 5 unless otherwise stated. To draw a vertical vein centered at the x-coordinate $p$, we use Dijkstra's algorithm to find the minimum distance path from ($p$, 0) to ($p$, 1) and penalize deviations in the x-direction from $p$. This procedure yields a vein that approximately follows the vertical line $x = p$, but has some inherent stray due to the disorder of the underlying network. For a network with densely packed veins, this procedure decreases the chance of spuriously overlapping veins. Lifting the penalty on deviation from $p$ would yield a vein that is inclined to follow natural curved paths in the underlying lattice, resulting in veins of smaller total length but with the cost of increased vein crossings. 

\section{Threshold Choice}\label{appendix:thresh}
\begin{figure}
\begin{center}
	\includegraphics[width=70mm]{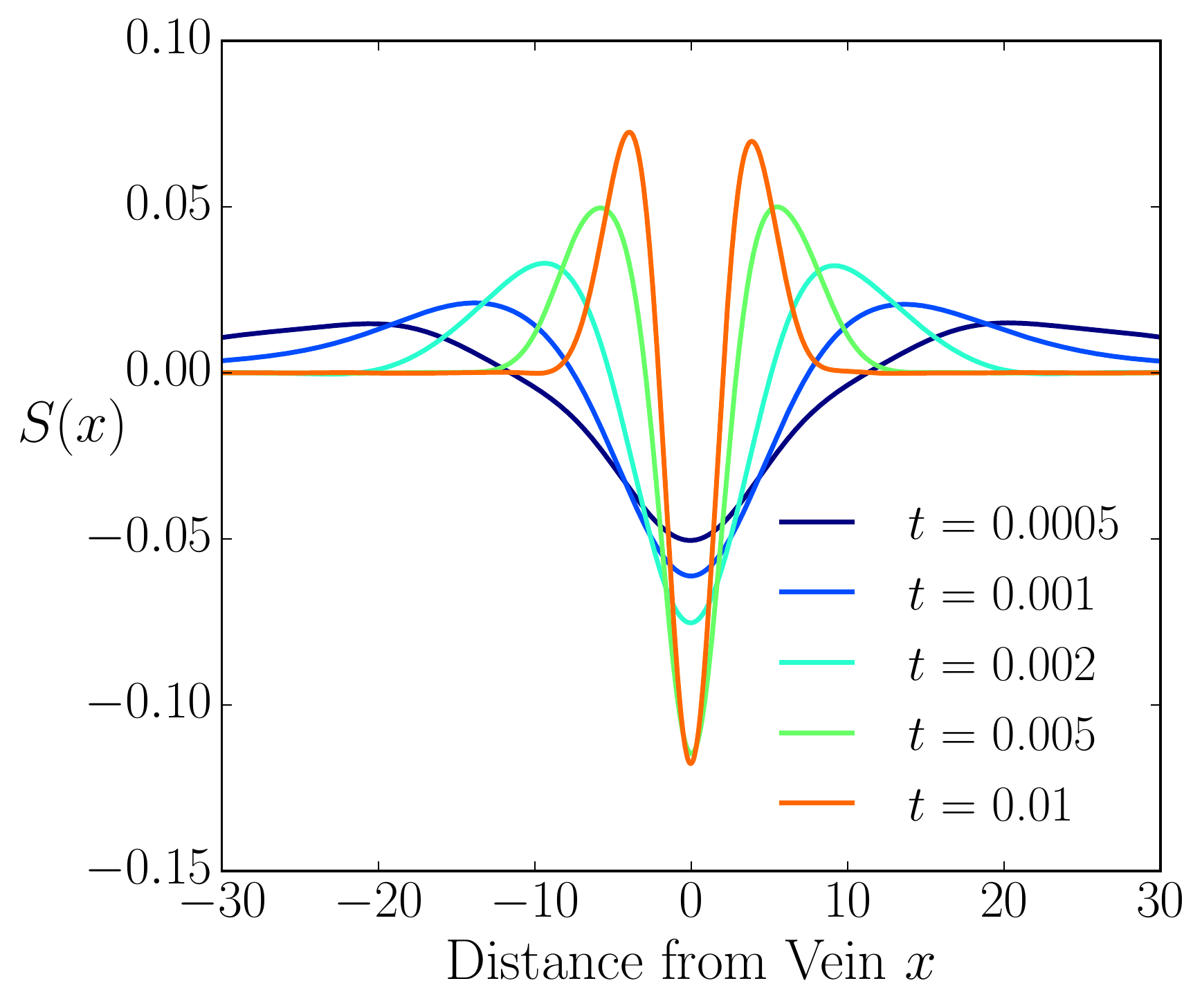} 
\caption{Comparison of $S(x)$ for different damage zone thresholds $t$ for a RP tiling with 5000 nodes with bulk edges of conductance 1 and a single central vein of conductance 5. As $t$ becomes smaller the shielding effect becomes more diffuse, but $S(x)$ maintains the same characteristic shape. The value $t = 0.005$ is used for all calculations.} 
\label{thresh}
\end{center}
\end{figure}

The characteristic size of the damage zone is set by the threshold $t$. Changing the $t$ corresponds to changing the sensitivity of the edges to displaced current. Increasing $t$ will increase all damage zones, and the effects of the vein extend to a further distance. A lower threshold limit is imposed by the requirement that the shielding effect falls zero at a distance shorter than the system length. Setting the threshold too high results in too few edges in the damage zone, and the effect is too local to quantitatively describe the system. We find that $t = 0.005$ is a suitable threshold for the a system size on the order of $5000$ nodes, and we use this value for all calculations. For this value of $t$, a typical damage zone for an edge is $\sim 1 \%$ of the total network size. Changing the threshold smoothly deforms the shape of the shielding effect, as seen in Fig.~\ref{thresh}. For smaller $t$, the shielding effect is shallow and diffuse. For larger $t$, $S(x)$ has sharper peaks and increases in magnitude but dies out fairly close to the vein. All $S(x)$ are qualitatively similar and the choice of $t$ should ultimately be a value that yields reasonable damage zone sizes compared to the system size.

\begin{figure}
\begin{center}
\begin{tabular}{cc}
	\includegraphics[height=40mm]{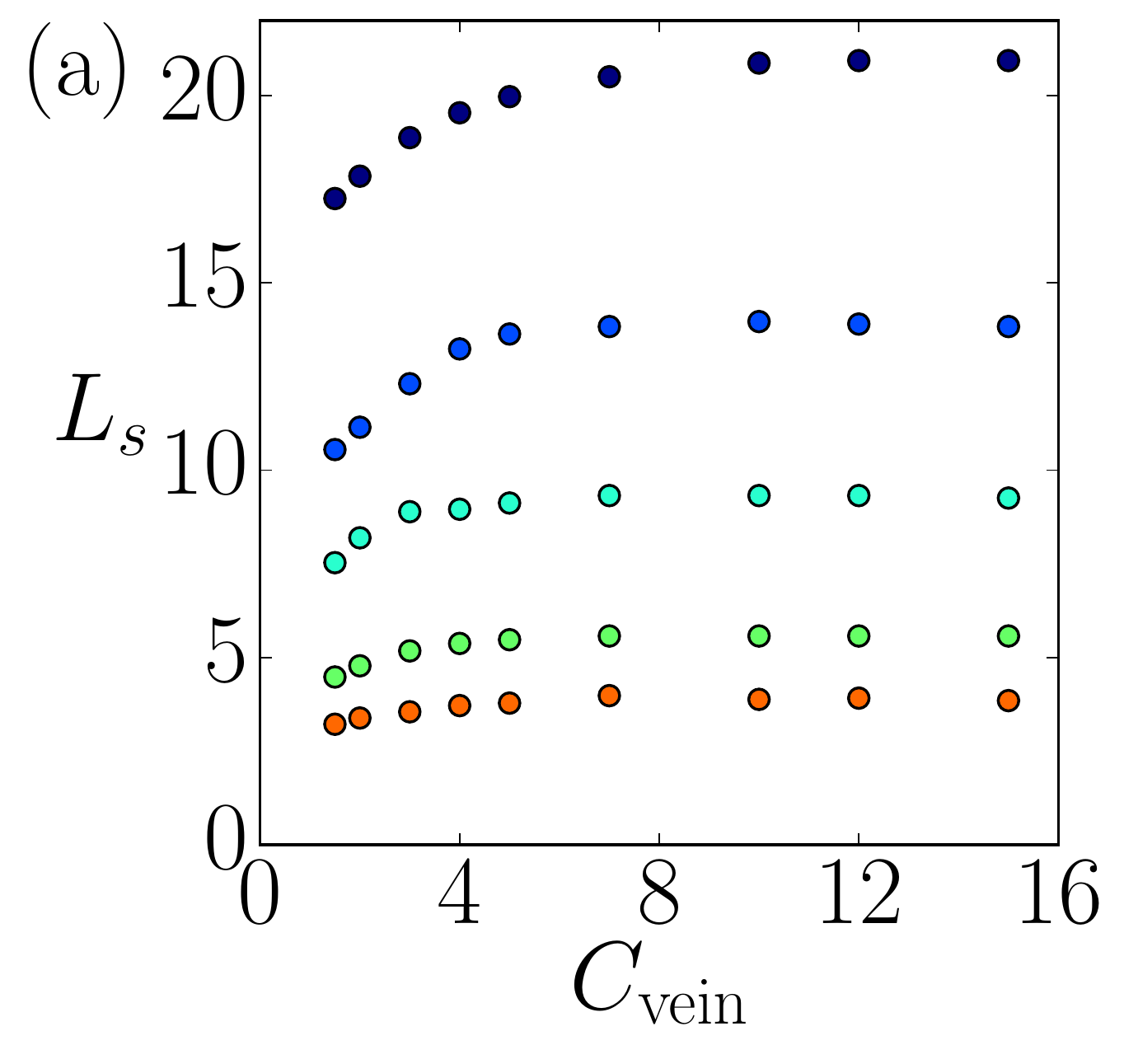}
	\includegraphics[height=40mm]{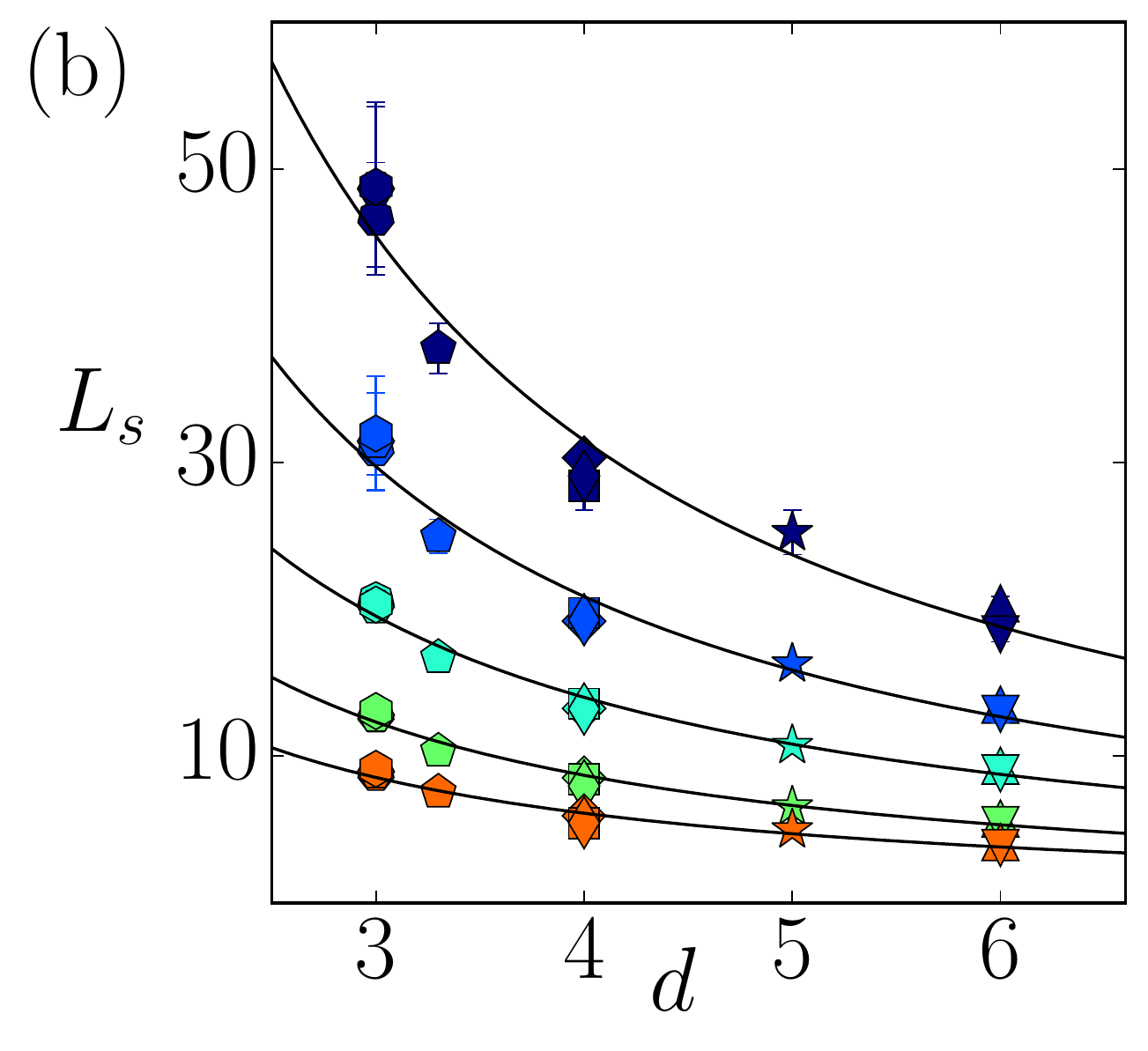}
\end{tabular}
\caption{Summary of our main results with varying values of $t$. We use $t = $ 0.0005, 0.001, 0.002, 0.005, and 0.01, with colors as in Fig. \ref{thresh}. (a)~For all values of $t$, $L_s$ becomes constant for $C_{vein} > 6$, although the asymptotic value increases with smaller $t$. (b)~For all values of $t$, the shielding length falls monotonically with increasing $d$, although the amplitude of the curves increases with smaller $t$.} 
\label{thresh2}
\end{center}
\end{figure}

Other results hold for a range of threshold values as well. Here we replicate two main results for the five values of $t$ shown in Fig.~\ref{thresh}. Figure~\ref{thresh2}~(a) shows that just as in Fig.~\ref{growVein}, the shielding effect emerges after $C_{vein}$ is larger than 1.5 times the bulk conductance. Once $C_{vein}$ exceeds around 5, $L_s$ becomes constant. This effect is not as prominent for $t=0.0005$, when the average damage zone becomes so large that finite network size effects come into play. Figure~\ref{thresh2}~(b) shows the same analysis as Fig.~\ref{SL_topology}(b). For all values of $t$, $L_s$ decreases monotonically as a function of $d$, and a best fit curve of the form $L_s \sim d^{\alpha}$ is drawn to guide the eye.

%

\section{Hierarchy in a Leaf Network} \label{appendix:real_networks}
\begin{figure}[h]
\begin{center}
\begin{tabular}{cc}
\includegraphics[width=45mm]{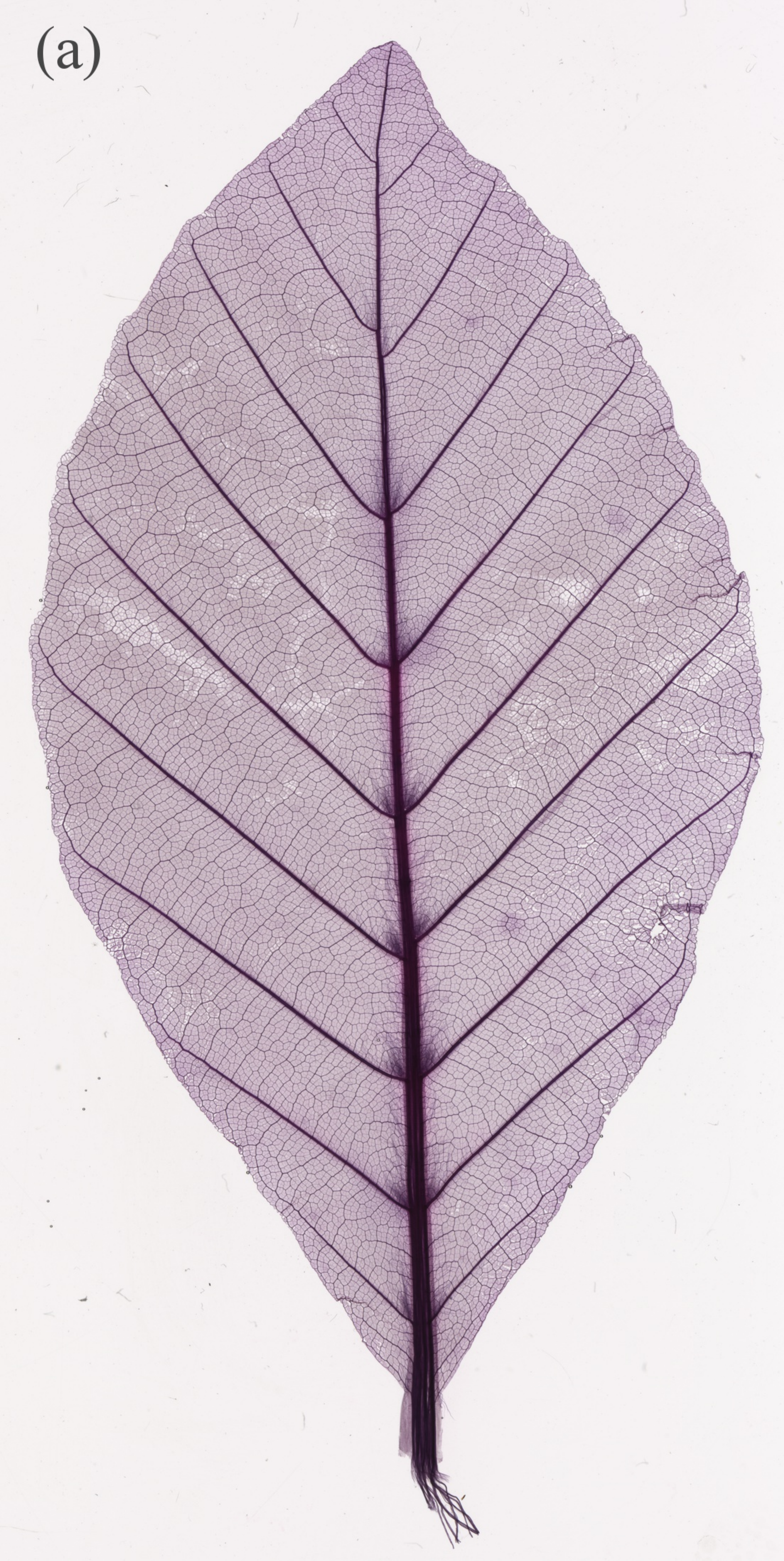} &
\includegraphics[angle=270, origin=t, width = 45mm]{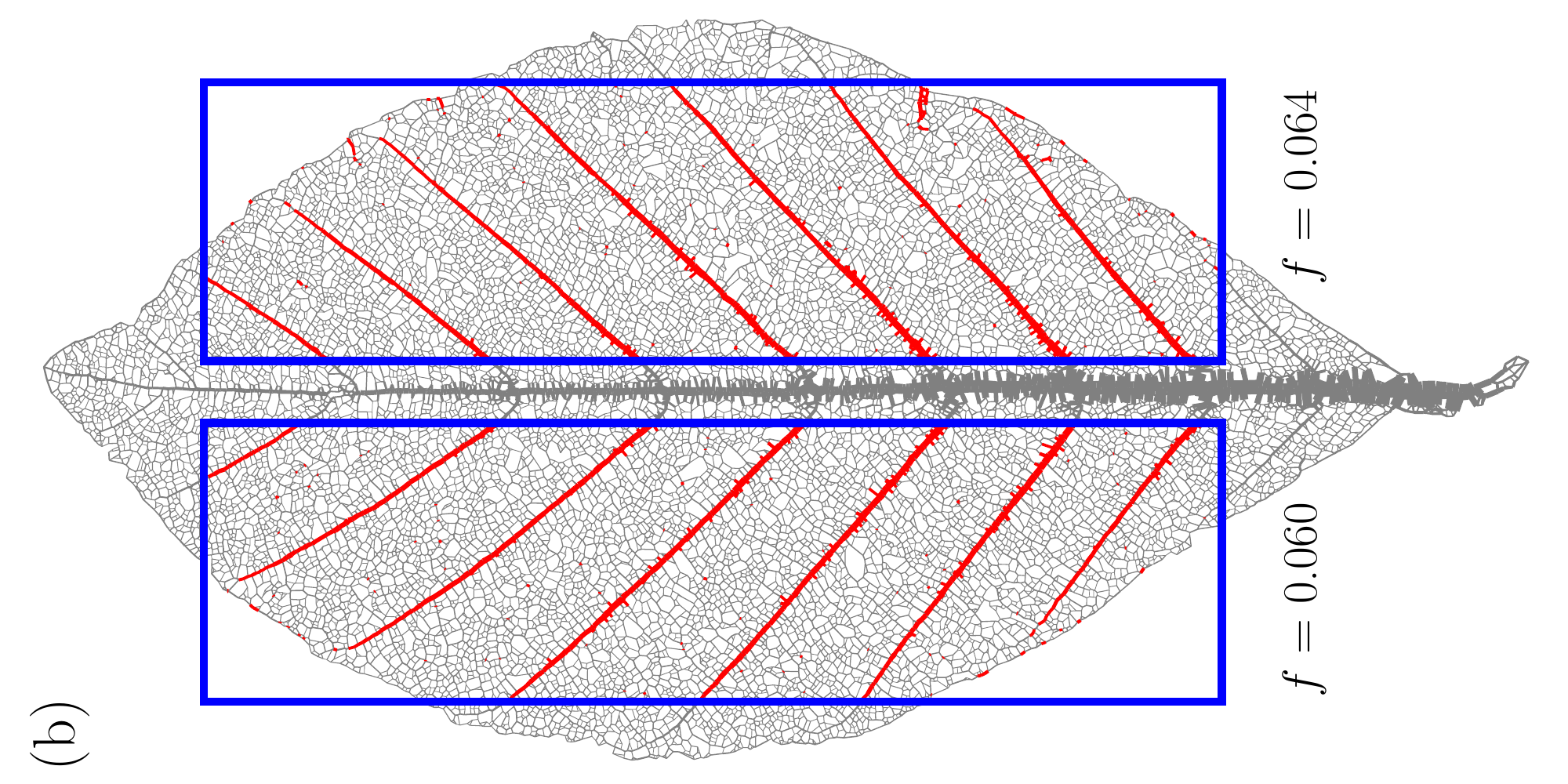} \\
\end{tabular}
\caption{Leaf venation of \textit{F. sylvatica}. (a)~The network of nodes and edges is extracted from a high-resolution scan. (b)~The occupation fraction is calculated for two rectangular sections of the leaf by computing the ratio of the number of high conductance vein edges (red) to the number of bulk edges (gray). For both sections, we find $f = 0.06$, which is reasonably close to the value that satisfies optimal resilience for a parallel vein hierarchy, $f = 0.10$.}
\label{leaf}
\end{center}
\end{figure}

To put the hierarchical networks considered in Section \ref{sec:multiple_veins} into context, we give an example of vein hierarchy in a leaf network. There are many diverse leaf vein patterns. While one might be hard-pressed to find a leaf with a grid hierarchy of veins, a structure of parallel secondary veins is quite common. To show that the optimal occupation fraction that we have derived in Fig. \ref{efficiency} is a reasonable value one might expect to see in a real world network, we calculate the occupation fraction for parallel veins in a leaf in Fig. \ref{leaf}. The leaf sample was bleached and stained to increase the visibility of the smallest veins. The venation network was extracted from a scan of the leaf using the Network Extraction Tool, providing node locations and edge weights \cite{Lasser2017}. We selected two regions of the leaf containing only the secondary veins forming the parallel network, shown by the blue boxes in Fig. \ref{leaf}. In these regions, any edge with a width more than one standard deviation above the mean width was considered to be a vein edge and colored in red, and all other edges were considered to be bulk edges. The occupation fraction $f$ was the ratio of vein edges to bulk edges, calculated to be $0.060$ for the left leaf section and $0.064$ for the right leaf section. For the parallel hierarchy we found f to be $0.10$, so for this single example the observed occupation fraction is close to the one that would yield optimal resilience.

\bibliography{damage}

\end{document}